\documentclass[apjl,ip,twocolumn]{aastex701}
\usepackage{comment}
\usepackage{ifthen}
\usepackage[hang,tight]{subfigure}
\usepackage{nicefrac}
\usepackage[export]{adjustbox}
\usepackage{mdframed}
\usepackage{colortbl}

\definecolor{LightGray}{rgb}{0.85,0.85,0.85}

\newcommand{\bestfitn}{\ensuremath{3.94 \pm 0.10}}

\newcommand{\bestfitbeta}
{\ensuremath{0.0552 \pm 0.0032}}

\newcommand{\bestfitMG}{\ensuremath{8.03 \pm 0.19}}

\newcommand{\bestfitMR}
{\ensuremath{8.21 \pm 0.16}}

\newcommand{\bestfitMgz}
{\ensuremath{9.56 \pm 0.18}}

\newcommand{\bestfitMrz}{\ensuremath{8.90 \pm 0.19}}

\newcommand{\bestfitMc}{\ensuremath{9.63 \pm 0.17}}

\newcommand{\bestfitMo}{\ensuremath{9.15 \pm 0.17}}

\newcommand{\bestfitMw}{\ensuremath{9.37\pm0.18}}

\newcommand{\bestfitMT}{\ensuremath{8.74\pm0.23}}

\newcommand{\bestfitMhg}{\ensuremath{9.57\pm0.17}}

\newcommand{\bestfitMhc}{\ensuremath{9.25\pm0.17}}

\newcommand{\bestfitMhr}{\ensuremath{8.86\pm0.17}}

\newcommand{\bestfitMho}{\ensuremath{8.72\pm0.17}}

\newcommand{\bestfitMhi}{\ensuremath{8.58\pm0.17}}

\newcommand{\bestfitMhz}{\ensuremath{8.51\pm0.17}}

\newcommand{\bestfitnhatpionly}{\ensuremath{5.167\pm0.095}}

\newcommand{\bestfitMGhatpionly}{\ensuremath{6.50\pm0.12}}

\newcommand{\extraerrorG}{\ensuremath{0.33}}

\newcommand{\extraerrorR}{\ensuremath{0.14}}

\newcommand{\extraerrorgz}{\ensuremath{0.060}}

\newcommand{\extraerrorrz}{\ensuremath{0.16}}

\newcommand{\extraerrorc}{\ensuremath{0.24}}

\newcommand{\extraerroro}{\ensuremath{0.13}}

\newcommand{\extraerrorw}{\ensuremath{0.17}}

\newcommand{\extraerrorT}{\ensuremath{0.30}}

\newcommand{\extraerrorhg}{\ensuremath{0.077}}

\newcommand{\extraerrorhc}{\ensuremath{0.068}}

\newcommand{\extraerrorhr}{\ensuremath{0.062}}

\newcommand{\extraerrorho}{\ensuremath{0.063}}

\newcommand{\extraerrorhi}{\ensuremath{0.059}}

\newcommand{\extraerrorhz}{\ensuremath{0.066}}

\newcommand{\extraerrorGhatpionly}{\ensuremath{0.095}}

\newcommand{\forloop}[5][1]%
{%
\setcounter{#2}{#3}%
\ifthenelse{#4}%
	{%
	#5%
	\addtocounter{#2}{#1}%
	\forloop[#1]{#2}{\value{#2}}{#4}{#5}%
	}%
	{%
	}%
}%

\newcommand{\ctbd}[1]{}

\newcommand{\C}{\ensuremath{^{\circ}C\;}}

\newcommand{\totNatlasobs}{\ensuremath{15,317}}

\newboolean{emulateapj}
\setboolean{emulateapj}{true}

\newboolean{astroph}
\setboolean{astroph}{true}

\shortauthors{Hartman et al.}
\shorttitle{HATPI Light Curve of 3I/ATLAS}
\ifthenelse{\boolean{emulateapj}}{
    
}{
    
}

\begin{document}

\title{
HATPI Pre-Perihelion Time-series Photometry of the Interstellar Comet 3I/ATLAS
}

\correspondingauthor{Joel D. Hartman}
\email{jhartman@astro.princeton.edu}

\author[0000-0001-8732-6166]{Joel D. Hartman}
\affil{Department of Astrophysical Sciences, Princeton University, NJ 08544, USA}
\email{jhartman@astro.princeton.edu}

\author[0000-0001-7204-6727]{G\'asp\'ar \'A. Bakos}
\affil{Department of Astrophysical Sciences, Princeton University, NJ 08544, USA}
\email{gbakos@astro.princeton.edu}

\author[0000-0002-5389-3944]{Andr\'es Jord\'an}
\affil{Facultad de Ingenier\'ia y Ciencias, Universidad Adolfo Ib\'a\~{n}ez, Av. Diagonal las Torres 2640, Pe\~{n}alol\'en, Santiago, Chile}
\affil{Departamento de Astronomía, Universidad de Chile, Casilla 36-D, Santiago, Chile}
\affil{Obstech/El Sauce Observatory, Coquimbo, Chile}
\email{ajordan@astrofisica.cl}

\author[0000-0001-7442-6926]{Sarah Thiele}
\affil{Department of Astrophysical Sciences, Princeton University, NJ 08544, USA}
\email{sarah.thiele@princeton.edu}

\author[0000-0002-8423-0510]{Zolt\'an Csubry}
\affil{Department of Astrophysical Sciences, Princeton University, NJ 08544, USA}
\email{zcsubry@astro.princeton.edu}

\author[0000-0003-4787-2335]{Geert Jan Talens}
\affil{Denys Wilkinson Building, Department of Physics, University of Oxford, OX1 3RH, UK}
\email{geertjan.talens@physics.ox.ac.uk}

\author[0000-0002-8585-4544]{Attila B\'odi}
\affil{Department of Astrophysical Sciences, Princeton University, NJ 08544, USA}
\email{abodi@princeton.edu}

\author[]{S\'andor Pigai}
\affil{Hungarian Astronomical Association}
\email{sandor.pigai@gmail.com}

\author[]{Istv\'an Domsa}
\affil{Hungarian Astronomical Association}
\email{istvan.domsa@gmail.com}

\author[0009-0007-3707-4846]{Anthony Keyes}
\affil{Department of Astrophysical Sciences, Princeton University, NJ 08544, USA}
\email{anthony.keyes@princeton.edu}

\author[]{Vincent Suc}
\affil{Facultad de Ingenier\'ia y Ciencias, Universidad Adolfo Ib\'a\~{n}ez, Av. Diagonal las Torres 2640, Pe\~{n}alol\'en, Santiago, Chile}
\affil{Obstech/El Sauce Observatory, Coquimbo, Chile}
\email{vincent.suc@gmail.com}

\author[0009-0000-7483-4495]{Adriana Gaitan}
\affil{Department of Astrophysical Sciences, Princeton University, NJ 08544, USA}
\email{ag2437@princeton.edu}

\author[0009-0004-3477-9064]{Antoine Thibault}
\affil{Observatoire astronomique de l’Université de Genève, Chemin Pegasi 51, CH-1290 Versoix, Switzerland}
\email{antoine.tbt@gmail.com}


\begin{abstract}
\setcounter{footnote}{10}
	HATPI is a recently commissioned time-domain facility at Las Campanas Observatory, Chile, that uses 64 wide-angle, 9.6\,cm diameter lenses and back-illuminated CCDs, yielding a mosaic field-of-view of 7,100 square arcdegrees, observing the night sky at a cadence of 45\,s and a spatial scale of 19\farcs 7\,pixel$^{-1}$. In this paper, we present moving object time-series photometry with this facility, focusing on the interstellar comet 3I/ATLAS. 3I/ATLAS was first robustly recovered by HATPI on the night of 2025 July 2 (one night after its discovery) at a {\em Gaia} $G$-band magnitude of $G = 17.796 \pm 0.082$\,mag ($\pm 0.030$\,mag systematic uncertainty). The comet then increased in brightness to $G = 14.071 \pm 0.073$\,mag $\pm 0.030$\,mag by 2025 Sep 13, after which it became unobservable by HATPI as it approached perihelion. Before 3I/ATLAS achieved a brightness of $G = 16.396 \pm 0.029$\,mag $\pm 0.030$\,mag on 2025 Aug 6, it could be detected when stacking all HATPI observations from a single night, while after this date it is sufficiently bright to detect in individual 45\,s exposures. We do not detect evidence for significant short-time-scale variations in the brightness of 3I/ATLAS after Aug 6. Compared to other light curves in the literature, the HATPI photometry exhibits a somewhat steeper rise in brightness with decreasing heliocentric distance, $r_{H}$. The HATPI magnitudes are well-fit as a power law function of $r_{H}$, with an exponential index of $n = \bestfitnhatpionly$, over the range $2.14$\,AU\,$ < r_{H} < 4.44$\,AU, compared to $n = \bestfitn$ when fitting together with other literature observations. We find that the phase function is constrained to $\beta = \bestfitbeta$\,mag\,deg$^{-1}$.
\setcounter{footnote}{0}
\end{abstract}



\section{Introduction}
\label{sec:introduction}

Comet 3I/ATLAS (initially dubbed C/2025 N1) was discovered by the ATLAS survey on 2025 July 1 \citep{denneau:2025}. The object has an extreme hyperbolic orbit, indicating an interstellar origin. It is the third such object discovered to date (the previous two being 1I/$^{\text{`}}$Oumuamua and 2I/Borisov, \citealp{meech:2017,borisov:2019}).

Given the rare and fleeting opportunity to study a solid body of interstellar origin, 3I/ATLAS has been the target of significant observational efforts, including imaging \citep{seligman:2025,jewitt:2025a,xing:2025,cordiner:2025,hoogendam:2025,qicheng:2025,scarmato:2025,roth:2025,lisse:2025b,serraricart:2025,combi:2025,scarmato:2025b,tan:2026,hui:2026}, time-series photometry \citep{seligman:2025,bolin:2025,kareta:2025,fuentemarcos:2025,chandler:2025,feinstein:2025,santanaros:2025,martinez:2025,beniyama:2025,salazarmanzano:2025,frincke:2025,tonry:2025,ye:2025,jewitt:2025,qicheng:2025,trigorodriguez:2025,eubanks:2025,lisse:2025b}, spectroscopy \citep{seligman:2025,opitom:2025,alvarezcandal:2025,belyakov:2025,kareta:2025,fuentemarcos:2025,yang:2025,santanaros:2025,puzia:2025,cordiner:2025,rahatgaonkar:2025,salazarmanzano:2025,hutsemekers:2025,coulson:2025,hoogendam:2025,trigorodriguez:2025,roth:2025,hinkle:2025,lisse:2025b,hoogendam:2025b,lisse:2026,hoogendam:2026,belyakov:2026}, time-domain radio observations \citep{sheikh:2025,jacobsonbell:2025}, and polarimetry \citep{gray:2025,choi:2026}. 

These observations have in turn inspired a number of theoretical investigations into the physical properties and Galactic origin of the object \citep{hopkins:2025,taylor:2025,hibberd:2025,guo:2025,perezcouto:2025,keto:2025a,cloete:2025,keto:2025b,yaginuma:2026,maggiolo:2026,neukart:2025,ahuja:2025,forbes:2026,scarmato:2025b,scarmato:2026,ahuja:2026}, and proposals for missions to encounter the object \citep{hibberd:2026} or to perform novel observations of it \citep{barbieri:2026}.

In this paper we contribute pre-perihelion time series photometry of 3I/ATLAS obtained by the Hungarian-made Automated Telescope PI steradians (HATPI) facility. HATPI is a high-cadence, moderately high-spatial resolution time-domain survey that uses 64 fast small telescopes in a mosaic configuration to monitor the full sky visible (above the horizon by at least approximately 35 degrees) from Las Campanas Observatory in Chile. Some of the key science aims of the HATPI facility are to discover long-period transiting giant planets, bright rapidly-evolving transients, and small asteroids moving quickly past the Earth. The orbit of 3I/ATLAS brought it across the southern sky during 2025 May--Sep as it approached perihelion. As a result, it was observed by HATPI. 

In the following Section we discuss the HATPI instrument, the observations of 3I/ATLAS, and our process for extracting time-series photometry for this moving object. In Section~\ref{sec:discussion} we analyze the light curve in the context of previously reported time-series observations of 3I/ATLAS, and use it to measure the heliocentric index of 3I/ATLAS through the HATPI band-pass. We conclude in Section~\ref{sec:conclusion}.

\section{Data Collection and Processing}\label{sec:instrumentandpipeline}

In this section we describe the HATPI instrument, the general data processing methods for the instrument that are relevant to the 3I/ATLAS observations, the observations of 3I/ATLAS, and the additional analysis processes applied specifically to the observations of this object to derive a light curve for it.

\subsection{The HATPI Instrument}\label{sec:instrument}

A thorough discussion of the HATPI instrument and data analysis pipeline will be provided in an upcoming paper (Bakos et al., in preparation), here we provide a brief summary.

The HATPI instrument is located at Las Campanas Observatory in Chile, which is managed by the Carnegie Institution for Science. The instrument uses 64 Mitakon 154\,mm f/1.6 lenses, each with a diameter of 96\,mm, together with 64 MicroLine ML230 cameras from the Finger Lakes Instrumentation company, each with a E2V CCD230-42 2048$\times$2048 back-side-illuminated detector, to observe a combined mosaic field-of-view of 7,055 square degrees (i.e., 17.1\% of the full celestial sphere, or $0.7\pi$\,steradians) at a plate scale of 19\farcs7\,pixel$^{-1}$. This massive field of view is observed at a cadence of 45\,s, with all 64 cameras observing synchronously. 

Observations are obtained through a custom broad-band filter that covers the wavelength range 430\,nm $< \lambda < 890$\,nm ($\lambda = 660$\,nm central wavelength). While not equivalent, it is fairly close to the {\em Gaia} $G$ band-pass which covers $\sim 330$\,nm to $\sim 1050$\,nm, or $\sim 400$\,nm to $\sim 860$\,nm at half maximum transmissivity. We use {\em Gaia} photometry to calibrate the photometry from HATPI, and report HATPI measurements as $G$-band measurements, though it should be noted that the HATPI bandpass is not identical to the $G$-bandpass.

At new moon, and in uncrowded regions of the sky, HATPI reaches a $5\sigma$ detection threshold of $G = 16.5$\,mag in a single exposure. Sources with $G \lesssim 9$\,mag are saturated in the HATPI images. 

The 64 lens+camera systems (referred to as instrument holder units, or IHUs) are all attached to a single massive equatorial-drive mount. The mount tracks the sky at the sidereal rate for one hour, before slewing back to its initial hour angle position, and tracking again. In addition to the general tracking that is provided by the mount, each IHU has 3 separate motors that allow for independent micro-tracking and focus corrections of that IHU. As a result, the pointing is stable at the sub-pixel level at the centers of each of the 64 individual fields forming the mosaic, and to the pixel level across the entire mosaic, over the course of each 1\,hr tracking period.

Note that because HATPI tracks at the sidereal rate, 3I/ATLAS moves slightly with respect to the camera pointing over the course of each 45\,s exposure. The maximum motion of 3I/ATLAS within a single exposure over the time window of the observations presented here is $\sim 1\farcs3$, which is less than $0.07$ pixels. Because this is less than $0.1$\,pixels, no special corrections are made for the slight elongation of 3I/ATLAS in the images compared to point sources.

\subsection{HATPI Image Processing}\label{sec:pipeline}

The images gathered by HATPI are automatically processed in real-time by server computers located in the HATPI enclosure at Las Campanas. The processing steps relevant to observations of moving objects include CCD image calibration, astrometry, and image subtraction. A MariaDB  database is used to keep track of the observations and data analysis products.

Standard CCD calibration steps, including over-scan correction, trimming, dark-current and bias correction, masking of bad pixels and columns, and flat-fielding, are carried out using tools in the {\sc FITSH} package \citep{pal:2012}. Master bias, dark, and sky-flat calibration images are obtained nightly (conditions permitting). Sky-flats are used to correct for small-scale variations in the sensitivity, while dome-flats (obtained using a hand-mountable luminescent flat-field screen) are used to correct for large-scale variations in sensitivity.

An astrometric solution is derived for each of the calibrated frames using the {\em Gaia} DR2 point-source catalog \citep{gaiadr2} as the astrometric reference. We use DR2 rather than the more recent DR3 because DR3 was not yet available at the time that the processing pipeline was developed. In order to maintain long-term stability of the data products, we intend to use DR2 for the duration of the HATPI project. Astrometric and photometric differences between DR2 and DR3 are negligible at the precision of HATPI. The {\em Gaia} catalog is matched to the sources detected in the HATPI images using a custom-built astrometry package that is based in part on the \url{astrometry.net} \citep{lang:2010} software and the {\sc FITSH} tools. The World Coordinate System (WCS) solution determined for each image is stored in a database to facilitate the quick identification of images covering a particular location on the sky. The median error in the astrometric solution for HATPI observations of 3I/ATLAS is $0\farcs95$.

Optimal image subtraction \citep{alard:1998} is performed for each calibrated frame using the {\sc fitrans} and {\sc ficonv} tools within the {\sc FITSH} package. The subtraction is performed against an empirical reference image constructed from $\sim 100$ observations of the same field, with the same IHU, obtained during clear weather conditions, within a few days of new moon. The calibrated image is spatially registered to the reference image using {\sc fitrans}, and then {\sc ficonv} is used to find an optimal transformation to apply to the reference image to match the background, flux scale and PSF (via convolution) of the registered calibrated image, before performing the subtraction. A discrete matrix, with components represented as polynomials in space over the image, is used to parameterize the convolution kernel.

\subsection{Observations of 3I/ATLAS}\label{sec:obs}

    \begin{deluxetable*}{lrrrrrrrrrr}
\tablewidth{0pc}
\tabletypesize{\scriptsize}
\tablecaption{
    Summary of HATPI observations of 3I/ATLAS.
    \label{tab:obsnights}
}
\tablehead{
    \multicolumn{1}{c}{Night} &
    \multicolumn{1}{c}{JD\tablenotemark{a}} &
    \multicolumn{1}{c}{RA\tablenotemark{a}} &
    \multicolumn{1}{c}{Dec\tablenotemark{a}} &
    \multicolumn{1}{c}{$r_{H}$\tablenotemark{a}\tablenotemark{b}} &
    \multicolumn{1}{c}{$\Delta$\tablenotemark{a}\tablenotemark{c}} &
    \multicolumn{1}{c}{$\alpha$\tablenotemark{a}\tablenotemark{d}} &
    \multicolumn{1}{c}{$\nu$\tablenotemark{a}\tablenotemark{e}} &
    \multicolumn{1}{c}
    {N. Obs} &
    \multicolumn{1}{c}
    {N. Clean Obs} &
    \multicolumn{1}{c}{N. IHUs} \\
    \multicolumn{1}{c}{} &
    \multicolumn{1}{c}{} &
    \multicolumn{1}{c}{(deg)} &
    \multicolumn{1}{c}{(deg)} &
    \multicolumn{1}{c}{(AU)} &
    \multicolumn{1}{c}{(AU)} &
    \multicolumn{1}{c}{(deg)} &
    \multicolumn{1}{c}{(deg)} &
    \multicolumn{1}{c}{} &
    \multicolumn{1}{c}{} &
    \multicolumn{1}{c}{}
}
\startdata
2025 May 1 & 2460797.8467 & 288.847363 & -18.718188 & 6.52 & 6.05 & 8.12 & 274.53 & 299 & 0 & 4 \\
2025 May 2 & 2460798.8453 & 288.743496 & -18.714663 & 6.49 & 6.00 & 8.09 & 274.60 & 303 & 303 & 4 \\
2025 May 3 & 2460799.8443 & 288.635249 & -18.711415 & 6.45 & 5.95 & 8.07 & 274.67 & 310 & 126 & 4 \\
2025 May 4 & 2460800.8425 & 288.522631 & -18.708477 & 6.42 & 5.90 & 8.04 & 274.75 & 227 & 0 & 4 \\
2025 May 5 & 2460801.8630 & 288.402761 & -18.705727 & 6.39 & 5.85 & 8.00 & 274.82 & 177 & 130 & 4 \\
2025 May 8 & 2460804.8393 & 288.025477 & -18.699409 & 6.29 & 5.70 & 7.87 & 275.05 & 244 & 244 & 5 \\
2025 May 9 & 2460805.8643 & 287.885565 & -18.697792 & 6.25 & 5.65 & 7.82 & 275.13 & 164 & 0 & 5 \\
2025 May 10 & 2460806.8156 & 287.751215 & -18.696554 & 6.22 & 5.60 & 7.77 & 275.21 & 306 & 105 & 6 \\
2025 May 11 & 2460807.8101 & 287.605774 & -18.695501 & 6.19 & 5.55 & 7.72 & 275.29 & 292 & 157 & 6 \\
2025 May 19 & 2460815.6971 & 286.264719 & -18.696346 & 5.92 & 5.18 & 7.12 & 275.95 & 43 & 43 & 1 \\
2025 May 20 & 2460816.7838 & 286.052017 & -18.697684 & 5.88 & 5.13 & 7.02 & 276.04 & 433 & 428 & 7 \\
2025 May 22 & 2460818.7781 & 285.643396 & -18.700733 & 5.81 & 5.03 & 6.81 & 276.22 & 450 & 109 & 7 \\
2025 May 23 & 2460819.7756 & 285.429831 & -18.702563 & 5.78 & 4.99 & 6.70 & 276.31 & 528 & 419 & 7 \\
2025 May 24 & 2460820.7744 & 285.209652 & -18.704589 & 5.75 & 4.94 & 6.59 & 276.40 & 564 & 548 & 7 \\
2025 May 25 & 2460821.6861 & 285.003334 & -18.706551 & 5.72 & 4.90 & 6.48 & 276.49 & 55 & 0 & 1 \\
2025 May 27 & 2460823.8288 & 284.496069 & -18.711798 & 5.64 & 4.80 & 6.21 & 276.69 & 601 & 4 & 7 \\
2025 May 28 & 2460824.8279 & 284.249291 & -18.714458 & 5.61 & 4.76 & 6.08 & 276.79 & 620 & 405 & 7 \\
2025 May 29 & 2460825.8612 & 283.986835 & -18.717325 & 5.58 & 4.71 & 5.93 & 276.89 & 111 & 111 & 3 \\
2025 May 30 & 2460826.8246 & 283.735712 & -18.720088 & 5.54 & 4.67 & 5.79 & 276.99 & 575 & 17 & 7 \\
2025 Jun 1 & 2460828.8270 & 283.192772 & -18.726059 & 5.48 & 4.58 & 5.49 & 277.19 & 252 & 0 & 4 \\
2025 Jun 5 & 2460832.7332 & 282.050760 & -18.737876 & 5.34 & 4.42 & 4.83 & 277.60 & 279 & 12 & 5 \\
2025 Jun 6 & 2460833.6968 & 281.751674 & -18.740703 & 5.31 & 4.38 & 4.66 & 277.70 & 253 & 36 & 6 \\
2025 Jun 8 & 2460835.6904 & 281.110289 & -18.746230 & 5.25 & 4.29 & 4.28 & 277.92 & 353 & 44 & 5 \\
2025 Jun 20 & 2460847.6254 & 276.606102 & -18.756166 & 4.84 & 3.84 & 1.68 & 279.36 & 452 & 191 & 5 \\
2025 Jun 21 & 2460848.6315 & 276.172358 & -18.753879 & 4.81 & 3.80 & 1.46 & 279.50 & 300 & 27 & 5 \\
2025 Jun 22 & 2460849.6795 & 275.711174 & -18.750773 & 4.78 & 3.76 & 1.27 & 279.64 & 202 & 29 & 2 \\
2025 Jun 23 & 2460850.7029 & 275.251888 & -18.746910 & 4.74 & 3.73 & 1.11 & 279.77 & 604 & 57 & 6 \\
2025 Jun 24 & 2460851.5864 & 274.848765 & -18.742849 & 4.71 & 3.70 & 1.03 & 279.90 & 10 & 0 & 1 \\
2025 Jun 26 & 2460853.8090 & 273.803436 & -18.729723 & 4.64 & 3.62 & 1.11 & 280.21 & 72 & 0 & 1 \\
2025 Jul 2 & 2460859.6355 & 270.869224 & -18.670638 & 4.44 & 3.44 & 2.58 & 281.08 & 580 & 281 & 8 \\
2025 Jul 3 & 2460860.6327 & 270.338913 & -18.656322 & 4.41 & 3.41 & 2.90 & 281.23 & 569 & 284 & 8 \\
2025 Jul 4 & 2460861.6911 & 269.767082 & -18.639606 & 4.37 & 3.38 & 3.25 & 281.40 & 502 & 381 & 8 \\
2025 Jul 6 & 2460863.5915 & 268.718920 & -18.605433 & 4.31 & 3.33 & 3.91 & 281.71 & 121 & 66 & 2 \\
2025 Jul 11 & 2460868.5703 & 265.841931 & -18.487985 & 4.15 & 3.20 & 5.74 & 282.56 & 446 & 235 & 7 \\
2025 Jul 15 & 2460872.5073 & 263.443857 & -18.362997 & 4.02 & 3.10 & 7.27 & 283.29 & 66 & 0 & 2 \\
2025 Jul 16 & 2460873.6755 & 262.712393 & -18.319991 & 3.98 & 3.08 & 7.74 & 283.51 & 406 & 93 & 6 \\
2025 Jul 17 & 2460874.5018 & 262.191372 & -18.287818 & 3.95 & 3.06 & 8.07 & 283.67 & 222 & 200 & 4 \\
2025 Jul 21 & 2460878.6074 & 259.543760 & -18.106427 & 3.82 & 2.98 & 9.74 & 284.51 & 439 & 162 & 6 \\
2025 Jul 23 & 2460880.5747 & 258.246933 & -18.006295 & 3.75 & 2.94 & 10.55 & 284.93 & 314 & 97 & 6 \\
2025 Jul 24 & 2460881.5169 & 257.620487 & -17.955244 & 3.72 & 2.92 & 10.94 & 285.14 & 144 & 36 & 3 \\
2025 Jul 25 & 2460882.5831 & 256.907046 & -17.895042 & 3.68 & 2.90 & 11.38 & 285.38 & 337 & 181 & 6 \\
2025 Aug 6 & 2460894.5949 & 248.703682 & -17.040573 & 3.29 & 2.74 & 16.20 & 288.41 & 382 & 301 & 5 \\
2025 Aug 7 & 2460895.5907 & 248.019423 & -16.955835 & 3.26 & 2.72 & 16.58 & 288.70 & 348 & 279 & 5 \\
2025 Aug 8 & 2460896.5490 & 247.361838 & -16.872492 & 3.23 & 2.71 & 16.93 & 288.98 & 203 & 201 & 4 \\
2025 Aug 11 & 2460899.5178 & 245.331763 & -16.603205 & 3.14 & 2.69 & 17.99 & 289.87 & 278 & 225 & 5 \\
2025 Aug 12 & 2460900.5732 & 244.613452 & -16.503637 & 3.10 & 2.68 & 18.36 & 290.21 & 116 & 88 & 4 \\
2025 Aug 13 & 2460901.5486 & 243.952204 & -16.410022 & 3.07 & 2.67 & 18.68 & 290.52 & 336 & 305 & 5 \\
2025 Aug 23 & 2460911.5309 & 237.357860 & -15.374258 & 2.76 & 2.61 & 21.49 & 294.14 & 161 & 161 & 3 \\
2025 Aug 25 & 2460913.5081 & 236.099581 & -15.155868 & 2.70 & 2.60 & 21.91 & 294.95 & 91 & 30 & 2 \\
2025 Sep 8 & 2460927.4864 & 227.751630 & -13.542803 & 2.28 & 2.55 & 23.17 & 301.93 & 41 & 40 & 2 \\
2025 Sep 9 & 2460928.4823 & 227.194063 & -13.425098 & 2.25 & 2.55 & 23.13 & 302.53 & 44 & 44 & 2 \\
2025 Sep 10 & 2460929.4807 & 226.639942 & -13.306873 & 2.22 & 2.55 & 23.07 & 303.14 & 40 & 39 & 2 \\
2025 Sep 11 & 2460930.4820 & 226.088892 & -13.188132 & 2.19 & 2.55 & 22.99 & 303.77 & 12 & 4 & 2 \\
2025 Sep 13 & 2460932.4796 & 225.003510 & -12.950752 & 2.14 & 2.54 & 22.77 & 305.08 & 17 & 16 & 1 \\
2025 Sep 14 & 2460933.4797 & 224.466869 & -12.831684 & 2.11 & 2.54 & 22.63 & 305.76 & 23 & 0 & 1 \\
\enddata 
\tablenotetext{a}{median value among observations from that night}
\tablenotetext{b}{Heliocentric distance of 3I/ATLAS.}
\tablenotetext{c}{Geocentric distance of 3I/ATLAS.}
\tablenotetext{d}{Sun-3I/ATLAS-Earth angle.}
\tablenotemark{e}{Apparent true anomaly angle of the target's heliocentric orbit position.}
    \end{deluxetable*}

In order to identify HATPI observations of the interstellar comet 3I/ATLAS, we first made use of the JPL Horizons Ephemeris Service (generated with SPICE kernel DE441, with the most recent 3I/ATLAS solution at the time of writing being JPL\#36) to determine the apparent RA/Dec position of 3I/ATLAS, as seen from Las Campanas Observatory, at 30 minute intervals spanning the time period 1 May 2025 through 30 Sep 2025. In practice, the query was performed multiple times over this period, and the full data analysis process described below was carried out as new observations continued to be gathered. Although the orbit of 3I/ATLAS has continued to be refined over time, its position was always known with significantly higher precision than the spatial resolution of HATPI.

Having obtained the ephemerides of 3I/ATLAS, we queried our database of WCS solutions to identify all images covering the position of 3I/ATLAS at the time of observation. Specifically, given position $({\rm RA}_{i}, {\rm Dec}_{i})$ at time $t_{i}$, we selected all images obtained at times between $(t_{i-1} + t_{i})/2$ and $(t_{i}+t_{i+1})/2$ containing position $({\rm RA}_{i}, {\rm Dec}_{i})$. We then perform a monotonic cubic spline interpolation \citep{steffen:1990} of the tabulated RA and Dec values to determine a more precise estimate of the position at the time of each observation, and use the WCS solution for the reference image to evaluate the X and Y pixel position of 3I/ATLAS on the subtracted images.

\begin{figure*}[t]
\includegraphics[width=0.99\textwidth]{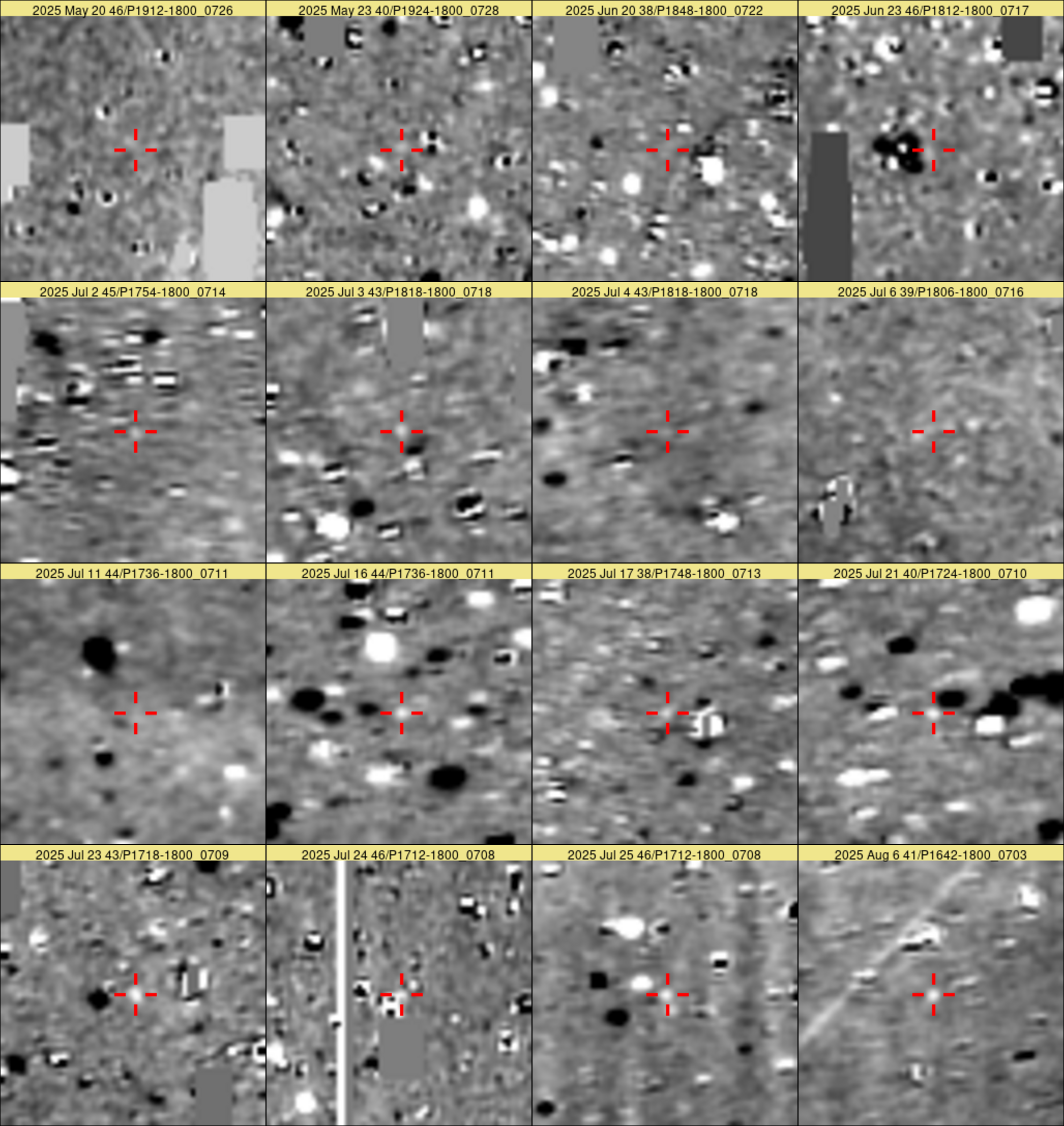}
\caption{
One hour stacked HATPI subtracted images of 3I/ATLAS from each of the nights on which 3I/ATLAS is formally detected with $3\sigma$ confidence in the full night binned light curve (nights corresponding to blue points in Fig.~\ref{fig:maglcnightly}). Individual 45\,s exposures that are flagged as problematic (Sec~\ref{sec:cleaning}) are excluded from the stacking. Each image stamp is $21\arcmin\times21\arcmin$, displayed with North up and East to the left. The red lines indicate the position of the nucleus of 3I/ATLAS in the center of each stamp. Grey rectangular regions are masked from the image subtraction process due to the presence of saturated stars. Other sources are variable objects, residual Poisson noise from bright stars, or other systematic residuals from poor subtractions (typically most notable for bright stars). Because 3I/ATLAS was at low Galactic latitude, the density of such objects is high in these stacked subtracted images. Individual 45\,s exposures are aligned on the position of 3I/ATLAS before stacking, causing background variable star sources (which can be either positive or negative) to show trails. For each night, we show a single IHU/Field as indicated in the labels after the date, while additional IHU/Fields may be available and included in the full night binned light curves. Note that the stacked images are for display purposes only. The photometry is performed on the individual 45\,s exposures, and binning is then performed on the photometric measurements rather than the images. The detections on May 20, May 23, June 20, June 23, July 6, July 11, and July 21 are likely affected by contamination from neighboring variable sources, and are excluded from the analysis. The first night on which we consider the recovery of 3I/ATLAS to be reliable is 2025 Jul 2. Additional stamps are shown in Fig.~\ref{fig:imgstamps2}.
\label{fig:imgstamps}
}
\end{figure*}

\begin{figure*}[t]
    \centering
\includegraphics[width=0.99\textwidth]{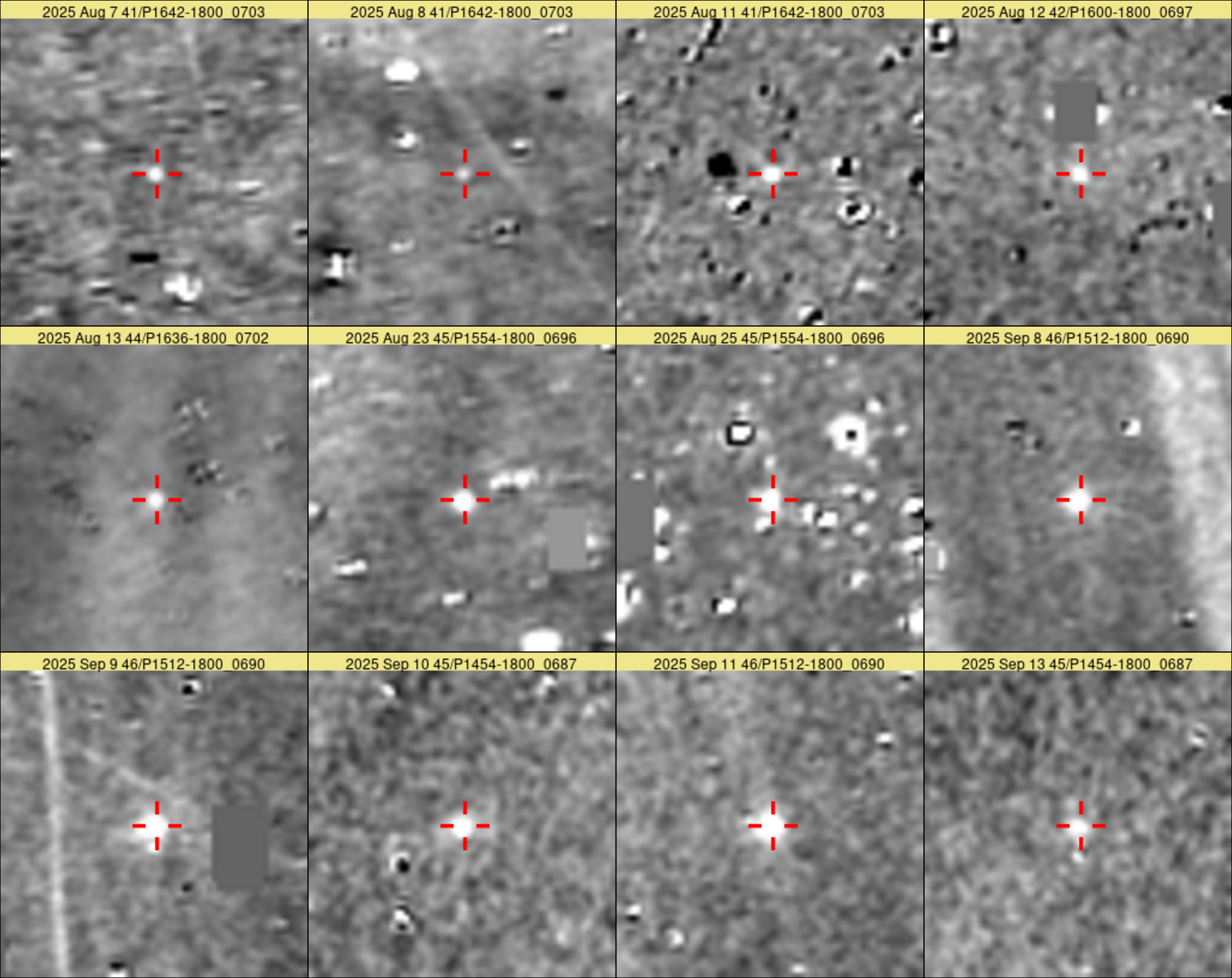}
\caption{
A continuation of Fig.~\ref{fig:imgstamps}. Here we show observations from Aug 7 through Sep 13.
\label{fig:imgstamps2}
}
\end{figure*}

We find that there are a total of \totNatlasobs\ individual HATPI images which cover the spatial position of 3I/ATLAS, and which were obtained between 2025 May 1 and 2025 Sep 13. A total of 6,238 of these images were collected after 2025 Jul 2, the date on which 3I/ATLAS first became detectable in the (stacked) HATPI images. Table~\ref{tab:obsnights} lists the nights on which HATPI observations of 3I/ATLAS are available. 3I/ATLAS was observed in total by 9 different HATPI IHUs, in 33 different distinct fields. The gaps in the observations are due to poor weather conditions at Las Campanas Observatory that prevented the HATPI system from operating. Unfortunately, the weather was unusually poor during July and August, resulting in longer-than-usual gaps in the data.

Figures~\ref{fig:imgstamps} and~\ref{fig:imgstamps2} show some of the 1\,hr stacked difference images of 3I/ATLAS from nights in which the comet was formally detected with $3\sigma$ confidence in the nightly binned HATPI light curve (Sec.~\ref{sec:photometry}).

\subsection{Photometry of 3I/ATLAS}\label{sec:photometry}

We used the {\sc fiphot} tool from the {\sc FITSH} package to perform forced aperture photometry of 3I/ATLAS on the subtracted images. For each image, photometry was performed at the RA/Dec position of the object on that image, determined from the JPL Horizons ephemeris. We performed the photometry using three fixed circular apertures of radii 1.45, 1.95 and 2.35 pixels, corresponding to angular radii of 28.6\arcsec, 38.4\arcsec, and 46.3\arcsec, respectively (photometry interpolated to a fixed $7 \times 10^{4}$\,km linear aperture is presented in Section~\ref{sec:discussion}). The {\sc fiphot} tool uses the transformation between the reference and calibrated images determined by {\sc ficonv} (Sec.~\ref{sec:pipeline}) to correct the photometry for changes in the flux scale and in the shape of the PSF. The resulting light curves are thereby corrected for atmospheric and instrumental variations, and can be effectively treated as ensemble-corrected relative flux measurements.

The flux values measured in ADUs are tied to the reference used for the image subtraction. There are a total of 33 references that contributed to the observations of 3I/ATLAS (one for each IHU/field combination in which 3I/ATLAS was observed), each on its own flux scale. To tie the photometry from the different references to a common scale we use {\em Gaia} DR2 $G$ and $BP-RP$ catalog photometry measurements. For each reference we find a polynomial transformation from the instrumental reference magnitudes, calculated directly from the reference fluxes, to the {\em Gaia} $G$ magnitude. The transformation depends on the $X$ and $Y$ positions of the sources, and their $BP-RP$ colors. We measure an average uncertainty of $0.030$\,mag on  this transformation, based on the standard deviation of the differences between the {\em Gaia} catalog $G$ magnitudes and the instrumental magnitudes transformed to the {\em Gaia} system for stars used in the fit. 

This same transformation can also be cast as a flux scaling factor $s$, such that $G = -2.5\log_{10}(sf)$, where $f$ is the flux measured on the reference system, and $G$ is the magnitude on the {\em Gaia} $G$-band system. We apply this scaling factor $s$ to the subtracted image flux measurements of 3I/ATLAS, assuming a value of $BP - RP = 1.0$\,mag for the color of the comet (slightly redder than the Solar value of $BP - RP = 0.85$\,mag, \citealp{andrae:2018}, based on early photometry of 3I/ATLAS indicating a color that is redder than the Sun). Note that the transformation does not have a sensitive dependence on color. Changing the assumed $BP - RP$ color by 0.15\,mag leads to a $< 0.02$\,mag systematic shift in the inferred $G$ magnitude, and a $0.003$\,mag standard deviation in the resulting light curve due to each IHU/Field having a slightly different color term. Although the color of 3I/ATLAS has evolved over time, we choose to fix the color in the transformation to {\em Gaia} so that the resulting light curve can be more directly interpreted as the light curve measured through the HATPI 430\,nm to 890\,nm bandpass.

\subsection{Cleaning Contaminated and Problematic Measurements}\label{sec:cleaning}

There are a number of effects that can produce artificial variations in the HATPI light curve of a moving object like 3I/ATLAS. Here we describe our methods for identifying affected observations, so that they may be filtered from the data, leaving a clean light curve for analysis.

To flag low quality images (e.g., such as those affected by clouds) that are likely to negatively impact the photometry for many objects, including 3I/ATLAS, we make use of subtracted image statistics that are calculated for all subtracted frames by the HATPI data analysis pipeline, and that are recorded in a database. We flag as problematic, subtracted images with a mean pixel value greater than $20$\,ADU or less than $-15$\,ADU, or with a standard deviation greater than $400$\,ADU, or a median absolute deviation greater than $100$\,ADU, or with a $5$ percentile value less than $-400$\,ADU, or a $95$ percentile value greater than $400$\,ADU. In addition to these cuts, we also impose a constraint on the mean of tiled medians statistic. This statistic is computed by splitting the subtracted frame into a grid of tiles, calculating the median of each tile, and then taking the mean of these different median values. We flag images where the statistic is greater than $20$\,ADU, or less than $-20$\,ADU. All of these cuts are chosen based on a visual inspection of the time-series values of the statistic over the observations containing 3I/ATLAS, with thresholds imposed where there is a clear deviation in the distribution of values between a long tail of outliers, and a clustering of points around a typical range. A total of 143 out of the 15,317 observations of 3I/ATLAS are flagged.

The majority of HATPI images contain satellite trails. Although only a small fraction of pixels are affected by satellite trails in a given image, in a collection of more than 10,000 images it is likely that a given target is impacted by satellite trails at least several times \citep[e.g.,][]{borlaff:2025}. We make use of a convolutional-neural-network method to identify pixels in the subtracted frames that are affected by satellite trails. The method will be described in an upcoming paper (Thiele, et al., in preparation). We use these identifications to flag any observations in the HATPI light curve where the photometric aperture contains one or more pixels contaminated by satellites. Altogether, we find that 10 out of the 15,317 observations of 3I/ATLAS are affected.

Due to the motion of 3I/ATLAS across the sky, coupled with the relatively low spatial resolution of the HATPI images, 3I/ATLAS is blended with background stars in a significant fraction of the HATPI observations. The image subtraction method corrects for this blending by subtracting the light of the star, as observed on the reference image. For non-variable stars, this method works reasonably well, however, there remains enhanced Poisson noise in the residual at the location of the star, which for bright stars, can overwhelm the signal from 3I/ATLAS. Additionally, for stars that vary in brightness, a positive or negative variable source may be present in the subtracted image. Blending with variable stars can thus produce artificial variations in the light curve of 3I/ATLAS. 

We include in the HATPI light curve a flag indicating observations where a source in the {\em Gaia} DR2 catalog with $G < 13$\,mag is within $1\arcmin$ of 3I/ATLAS. A total of 7,412 observations are flagged as such. 

To identify those sources that are variable, we made use of any HATPI light curves that are available for them. We also made use of the ASAS-SN Catalog of Variable Stars \citep{jayasinghe:2019}\footnote{Obtained from the ASAS-SN Sky Patrol website \url{https://asas-sn.osu.edu/variables} on 2026 Feb 1.}. The HATPI light curves are generally from 2024, though some stars also have 2022 and 2023 light curves available as well.  We calculated the Generalized Lomb-Scargle periodogram \citep{zechmeister:2009} using the implementation in {\sc VARTOOLS} \citep{hartman:2016:vartools}, marking any source with a formal $log_{10}$ false alarm probability less than $-50$ as a potential variable star. A total of 233 of the 1265 stars with HATPI light curves that were within 1\arcmin\ of 3I/ATLAS are flagged as potentially variable. We also found 29 stars that are listed as variable in the ASAS-SN catalog, and were within 1\arcmin\ of 3I/ATLAS in the HATPI observations. Three of the ASAS-SN variables lacked HATPI light curves, and two have HATPI light curves, but were not flagged as variable. The remaining 24 ASAS-SN variables were also identified as variables based on their HATPI light curves. An additional flag is added to the HATPI light curve of 3I/ATLAS to note observations where the object was blended with a star identified as potentially variable in this manner. A total of 3,418 observations (out of 15,317 total observations) are flagged as being blended with an identified variable source.

Altogether, combining the flags for problematic images, satellites, contamination from bright neighbors, and contamination from variable stars, we find that 8,023 of the observations are flagged as being impacted by one or more of these effects, leaving 7,294 remaining clean observations.

\subsection{Photometric Uncertainties}\label{sec:errors}

The photometric uncertainties returned by {\sc fiphot} already account for shot noise from the source and the sky background. However, because of the motion of 3I/ATLAS across the sky, changes in the contamination from neighboring sources contribute additional scatter to the photometric measurements. We account for this by imposing a floor on the individual photometric uncertainties. For each measurement, we take the uncertainty to be the maximum of the formal uncertainty from {\sc fiphot}, $1.483$ times the median absolute deviation (MAD) of all photometric measurements obtained before 2025 June 28, or $1.483$ times the MAD of all measurements obtained on the same night, and with the same IHU/Field as the target observation.  This process is performed separately for each of the three different photometric apertures used.

\subsection{Light Curve of 3I/ATLAS}\label{sec:lc}

\begin{figure*}[t]
\begin{center}
\subfigure[
    	\label{fig:fluxlcunbinnedbadtypes}]
        {\includegraphics[width=2\columnwidth]{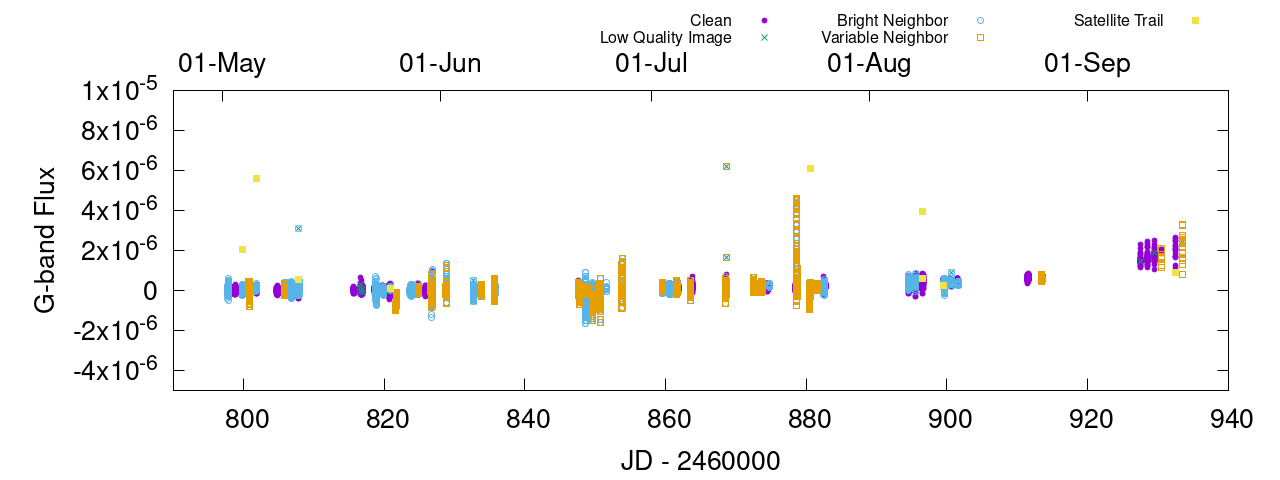}}
\subfigure[
    	\label{fig:fluxlcunbinnedcleaned}]
        {\includegraphics[width=2\columnwidth]{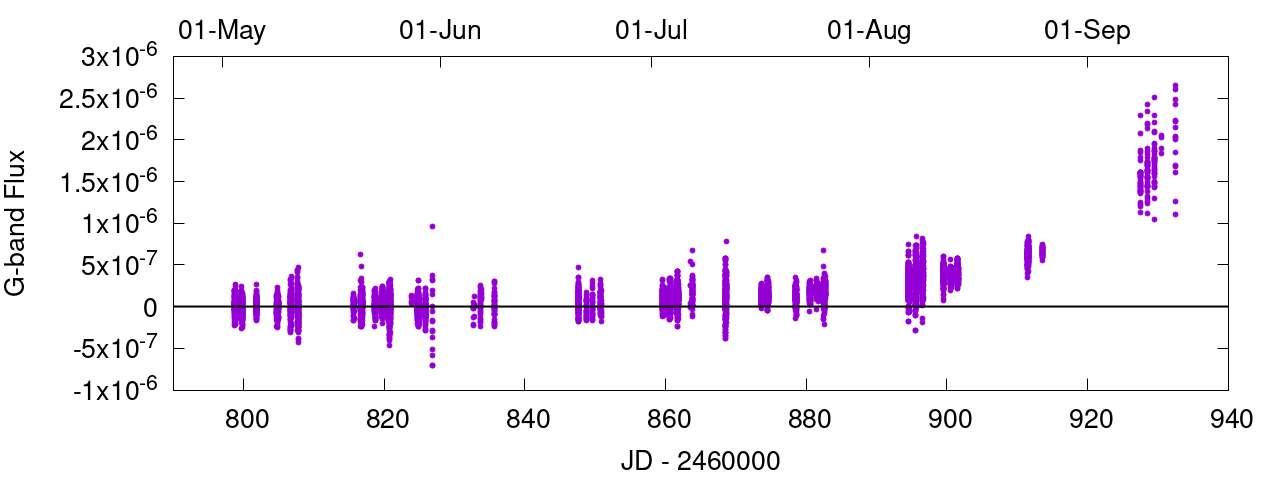}}
\caption{
Full, un-binned, HATPI light curve of 3I/ATLAS shown as fluxes $F$. These are calibrated to the {\em Gaia} $G$-band system such that the {\em Gaia} magnitude is equal to $-2.5\log_{10}(F)$. {\em Top:} all points are shown, including those that are rejected by the cleaning methods described in Section~\ref{sec:cleaning}. The color coding shows points with different quality flags. {\em Bottom:} same as above, but now including only the "clean" points from the top panel.
\label{fig:3iatlaslc}
}
\end{center}
\end{figure*}

\begin{splitdeluxetable*}{rllrrrrBrrrrcccc}
\tablewidth{0pc}
\tabletypesize{\scriptsize}
\tablecaption{
    Full Un-binned HATPI Light Curve of 3I/ATLAS.
    \label{tab:fullfluxlc}
}
\tablehead{
    \multicolumn{1}{c}{JD} &
    \multicolumn{1}{c}{IHU} &
    \multicolumn{1}{c}{Field} &
    \multicolumn{1}{c}{RA} &
    \multicolumn{1}{c}{Dec} &
    \multicolumn{1}{c}{Flux$_{0}$\tablenotemark{a}} &
    \multicolumn{1}{c}{eFlux$_{0}$} &
    \multicolumn{1}{c}
    {Flux$_{1}$\tablenotemark{b}} &
    \multicolumn{1}{c}
    {eFlux$_{1}$} &
    \multicolumn{1}{c}{Flux$_{2}$\tablenotemark{c}} &
    \multicolumn{1}{c}{eFlux$_{2}$} &
    \multicolumn{1}{c}{Bad Stats\tablenotemark{d}} &
    \multicolumn{1}{c}{Satellite\tablenotemark{e}} &
    \multicolumn{1}{c}{BrightNbr\tablenotemark{f}} &
    \multicolumn{1}{c}{VarNbr\tablenotemark{g}} \\
    \multicolumn{1}{c}{(d)} &
    \multicolumn{1}{c}{} &
    \multicolumn{1}{c}{} &
    \multicolumn{1}{c}{(deg)} &
    \multicolumn{1}{c}{(deg)} &
    \multicolumn{1}{c}{(counts)} &
    \multicolumn{1}{c}{(counts)} &
    \multicolumn{1}{c}{(counts)} &
    \multicolumn{1}{c}{(counts)} &
    \multicolumn{1}{c}{(counts)} &
    \multicolumn{1}{c}{(counts)} &
    \multicolumn{1}{c}{} &
    \multicolumn{1}{c}{} &
    \multicolumn{1}{c}{} &
    \multicolumn{1}{c}{}
}
\startdata
2460797.77285 & 39 & P1906-1800\_0725 & 288.8550012 & -18.7184374 & $-1.094e-07$ & $1.2e-07$ & $-6.508e-10$ & $1.3e-07$ & $5.408e-08$ & $1.5e-07$ & 0 & 0 & 1 & 0 \\
2460797.77334 & 39 & P1906-1800\_0725 & 288.8549511 & -18.7184355 & $2.060e-08$ & $1.2e-07$ & $4.233e-08$ & $1.3e-07$ & $8.416e-08$ & $1.5e-07$ & 0 & 0 & 1 & 0 \\
2460797.77382 & 39 & P1906-1800\_0725 & 288.8549014 & -18.7184337 & $-5.273e-08$ & $1.2e-07$ & $-2.495e-08$ & $1.3e-07$ & $-2.375e-08$ & $1.5e-07$ & 0 & 0 & 1 & 0 \\
2460797.77431 & 39 & P1906-1800\_0725 & 288.8548522 & -18.7184318 & $3.215e-08$ & $1.2e-07$ & $7.549e-08$ & $1.3e-07$ & $6.092e-08$ & $1.5e-07$ & 0 & 0 & 1 & 0 \\
2460797.77480 & 39 & P1906-1800\_0725 & 288.8548020 & -18.7184299 & $-1.306e-07$ & $1.2e-07$ & $-1.280e-07$ & $1.3e-07$ & $-1.649e-07$ & $1.5e-07$ & 0 & 0 & 1 & 0 \\
2460797.77529 & 39 & P1906-1800\_0725 & 288.8547520 & -18.7184280 & $-4.254e-07$ & $1.2e-07$ & $-3.856e-07$ & $1.3e-07$ & $-3.915e-07$ & $1.5e-07$ & 0 & 0 & 1 & 0 \\
2460797.77577 & 39 & P1906-1800\_0725 & 288.8547025 & -18.7184260 & $-1.353e-07$ & $1.2e-07$ & $-1.034e-07$ & $1.3e-07$ & $-1.122e-07$ & $1.5e-07$ & 0 & 0 & 1 & 0 \\
2460797.77626 & 39 & P1906-1800\_0725 & 288.8546530 & -18.7184240 & $-1.458e-07$ & $1.2e-07$ & $-2.719e-08$ & $1.3e-07$ & $4.196e-08$ & $1.5e-07$ & 0 & 0 & 1 & 0 \\
2460797.77675 & 39 & P1906-1800\_0725 & 288.8546033 & -18.7184220 & $-1.622e-07$ & $1.2e-07$ & $-1.469e-07$ & $1.3e-07$ & $-1.556e-07$ & $1.5e-07$ & 0 & 0 & 1 & 0 \\
2460797.77724 & 39 & P1906-1800\_0725 & 288.8545534 & -18.7184200 & $-1.068e-07$ & $1.2e-07$ & $-9.740e-08$ & $1.3e-07$ & $-1.254e-07$ & $1.5e-07$ & 0 & 0 & 1 & 0 \\
\enddata
\tablecomments{This table is published in its entirety in the machine-readable format.
      A portion is shown here for guidance regarding its form and content.}
\tablenotetext{a}{Flux through 28.6\arcsec\ aperture scaled so that Gaia-band magnitude $G = -2.5\log_{10}(F)$.}
\tablenotetext{b}{Flux through 38.4\arcsec\ aperture scaled so that Gaia-band magnitude $G = -2.5\log_{10}(F)$.}
\tablenotetext{c}{Flux through 46.3\arcsec\ aperture scaled so that Gaia-band magnitude $G = -2.5\log_{10}(F)$.}
\tablenotetext{d}{$1$ - bad image statistics; $0$ - observation ok.}
\tablenotetext{e}{$1$ - satellite trail; $0$ - no trail.}
\tablenotetext{f}{$1$ - contaminated by bright neighbor; $0$ - not contaminated.}
\tablenotetext{g}{$1$ - contaminated by variable neighbor; $0$ - not contaminated.}
\end{splitdeluxetable*}

Fig.~\ref{fig:3iatlaslc} shows the full un-binned flux light curve of 3I/ATLAS before and after cleaning. The data are available in Table~\ref{tab:fullfluxlc}. 

\begin{figure*}[t]
\begin{center}
\subfigure[
    	\label{fig:fluxlc1hr}]
        {\includegraphics[width=2\columnwidth]{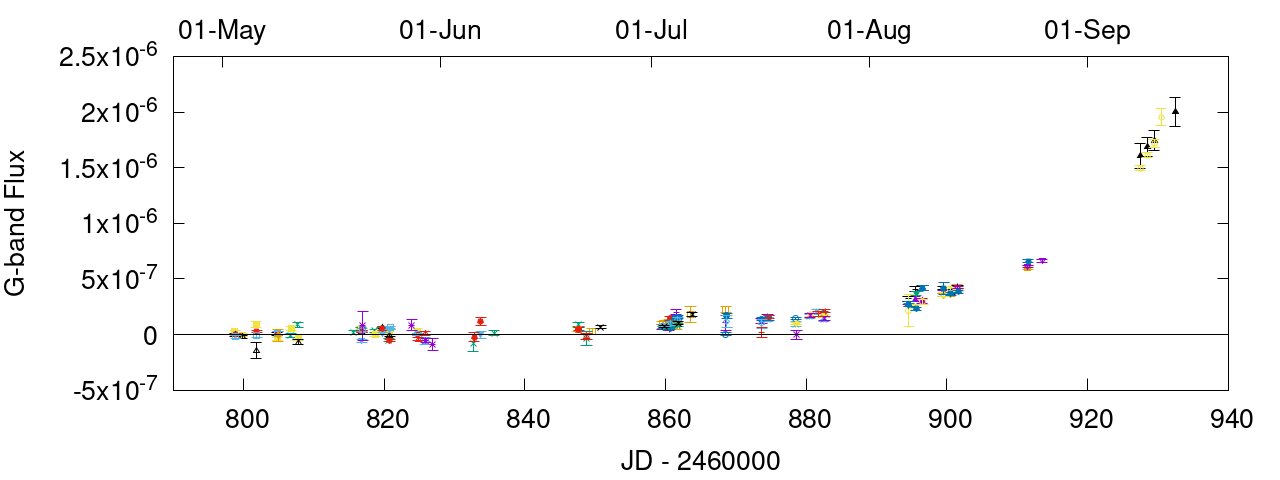}}
\subfigure[
    	\label{fig:fluxlcnightly}]
        {\includegraphics[width=2\columnwidth]{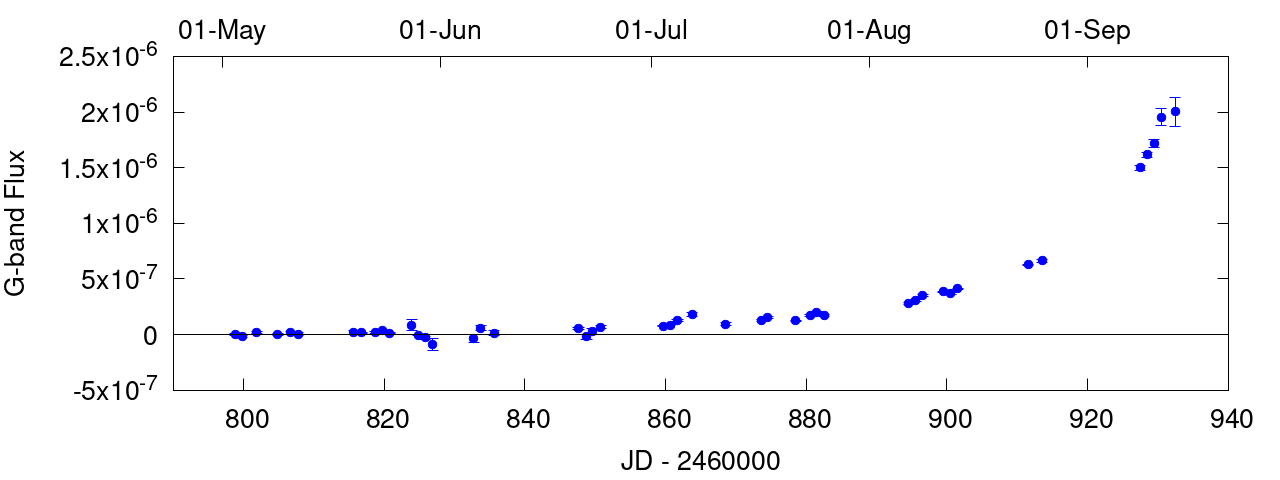}}
\subfigure[
    	\label{fig:maglcnightly}]
        {\includegraphics[width=2\columnwidth]{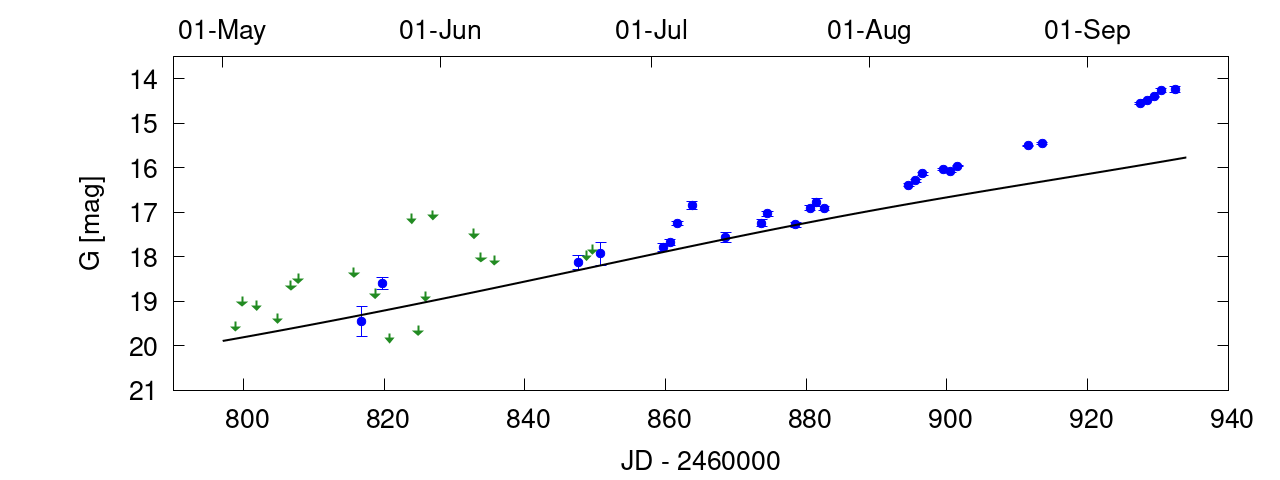}}
\caption{
{\em Top:} clean HATPI light curve of 3I/ATLAS shown as fluxes $F$ on the {\em Gaia} $G$-band system such that the {\em Gaia} magnitude is equal to $-2.5\log_{10}(F)$. We bin all observations obtained on a given night by a given IHU/Field using the weighted mean. Each point thus represents up to one hour of observations. Different colors indicate the different IHU/Fields. {\em Middle:} same as above, but now binning together all observations obtained on a given night. Each point in this case represents up to $\sim 6$ hours of observations. {\em Bottom:} same as {\em middle}, but showing the nightly binned light curve converted to magnitudes. Circles show points for which a non-zero flux is measured with at least $3\sigma$ confidence, while arrows indicate $3\sigma$ upper-limits on the brightness. The solid line shows the expected $T$ magnitude from JPL/Horizons.
\label{fig:binnedlc}
}
\end{center}
\end{figure*}

\begin{figure*}[t]
\begin{center}
\includegraphics[width=2\columnwidth]{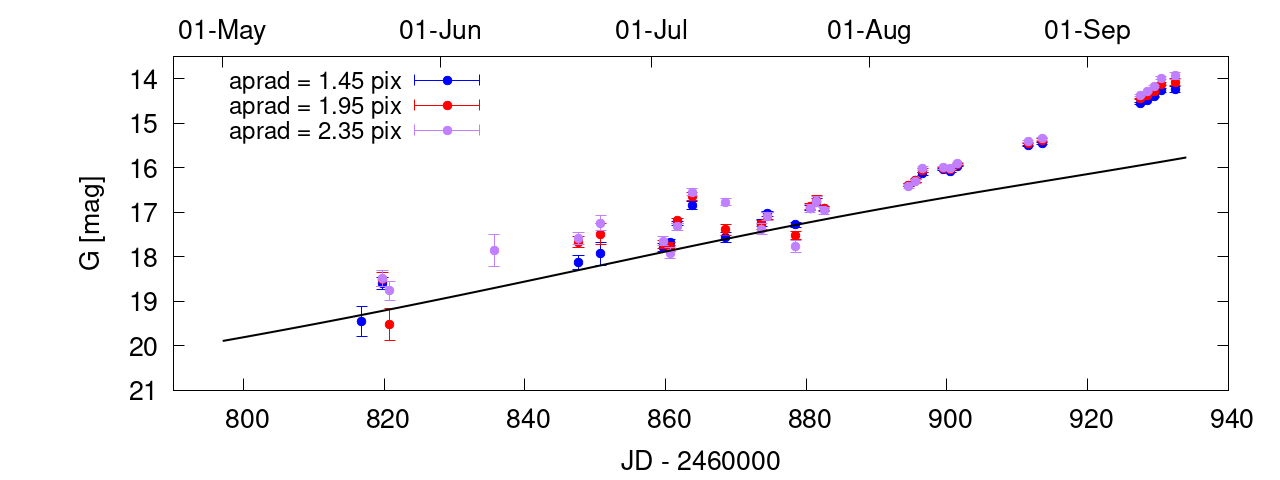}
\caption{
The nightly binned HATPI light curve of 3I/ATLAS for three different concentric, circular apertures. Here we omit upper-limit measurements. As a result, on some nights fewer than 3 apertures are displayed. All other figures show the 1.45 pixel aperture measurements. Large inconsistencies between the different apertures for some of the early-time observations indicate measurements that may be contaminated by blending with residual variable sources. At late times, as the cometary tail becomes resolved by HATPI, the largest aperture measurements are systematically brighter than the smaller aperture measurements.
\label{fig:maglcnightlycompaps}
}
\end{center}
\end{figure*}

Fig.~\ref{fig:binnedlc} shows the cleaned light curve binned in time using a weighted mean. We show the result when binning all observations for a given IHU/Field on a given night (i.e., an effective $1$\,hr binning), and when binning together all observations on a given night. For the light curve shown in magnitudes, we indicate the $3\sigma$ upper-limit on the brightness when the measured differential flux is within $3\sigma$ of zero. We also plot in this figure the expected $T$ magnitude of 3I/ATLAS from JPL/Horizons. This is computed using an expression equivalent to Eq.~\ref{eqn:mf} in Section~\ref{sec:discussion}, with $M_{F} = M_{V}$, $m = 2$, and $\Phi_{\alpha} \equiv 1.0$. JPL/Horizons adopts $M_{V} = 12.5$ and $n = 4.5$ for 3I/ATLAS. As seen in Fig.~\ref{fig:binnedlc}, these parameters do not provide a tight fit to the HATPI observations. In Section~\ref{sec:discussion} we fit eq.~\ref{eqn:mf} to the HATPI and literature light curves of 3I/ATLAS to obtain best-fit parameters.

The full-night binned light curve shows two potential early detections of 3I/ATLAS on the nights of 2025 May 20 and 2025 May 23, i.e., prior to the discovery, where we measure $G = 19.46 \pm 0.33$\,mag, and $G = 18.61 \pm 0.14$\,mag, respectively. However, we caution that 3I/ATLAS was in a crowded region of the sky at this time (Galactic latitude $b \approx -11^{\circ}$), and blending with other variable sources may have contaminated the measurements on these nights. It is worth noting that we also measure a negative flux on the night of 2025 May 29 that is $3\sigma$ less than $0$. This indicates that some variability contaminated measurements remain despite our efforts to flag and remove affected observations (Section~\ref{sec:cleaning}). Based on the model fit to the combined HATPI and published light curves of 3I/ATLAS (Section~\ref{sec:discussion}) we find that the expected magnitude of 3I/ATLAS on 2025 May 20 is $G = 19.55$\,mag, while on 2025 May 23 it is $G = 19.40$\,mag. The potential detection on 2025 May 20 is thus within $1\sigma$ of its expected value, while the possible 2025 May 23 observation would be $5.6\sigma$ brighter than its expected value. Given the overall uncertainty associated with these early measurements we exclude them from further analysis.

As a further check on potential variability blending, Fig.~\ref{fig:maglcnightlycompaps} compares the nightly binned light curve, in magnitudes, from the three different apertures used for photometry. On some of the earlier nights in the light curve there are large $> 1\sigma$ differences in the binned magnitudes between the different apertures. This is a signal of possible contamination from nearby sources. Nights of particular concern include June 20, June 23, July 6, July 11, and July 21. We exclude from further analysis these observations, together with the two early observations in May where the flux was formally detected with greater than $3\sigma$ confidence. 

The first night on which 3I/ATLAS appears to be robustly detected, without additional caution regarding potential contamination, is the night of July 2. On this night we detected a positive flux for 3I/ATLAS with $13\sigma$ confidence, and measure $G = 17.796 \pm 0.082$\,mag.

Fig.~\ref{fig:maglcnightlycompaps} also shows systematically brighter measurements for larger apertures during later nights in the light curve (i.e., following 2025 Sep 8). We attribute this to the cometary tail of 3I/ATLAS extending past the largest aperture on these dates.

\begin{figure*}[t]
\begin{center}
\subfigure[
    	\label{fig:unbinmaglc1}]{\includegraphics[width=0.32\textwidth]{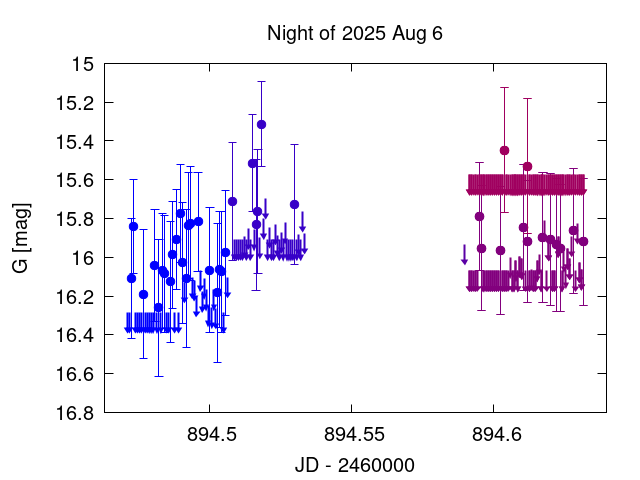}}
\subfigure[
    	\label{fig:unbinmaglc2}]{\includegraphics[width=0.32\textwidth]{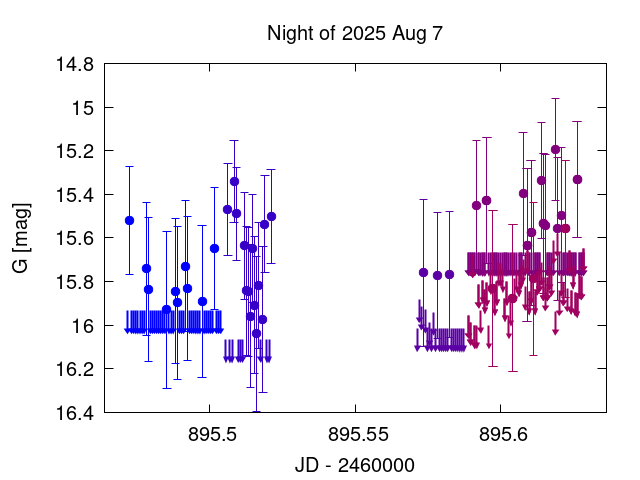}}
\subfigure[
    	\label{fig:unbinmaglc3}]{\includegraphics[width=0.32\textwidth]{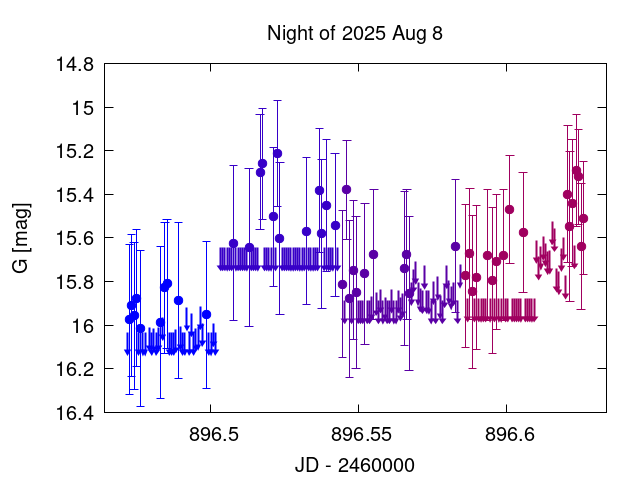}}
\subfigure[
    	\label{fig:unbinmaglc4}]{\includegraphics[width=0.32\textwidth]{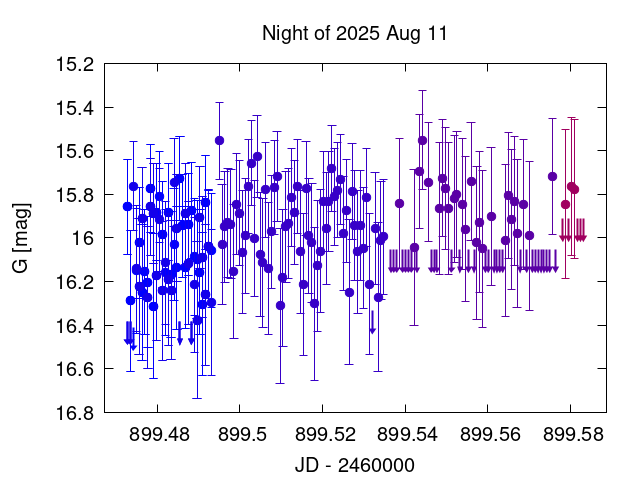}}
\subfigure[
    	\label{fig:unbinmaglc5}]{\includegraphics[width=0.32\textwidth]{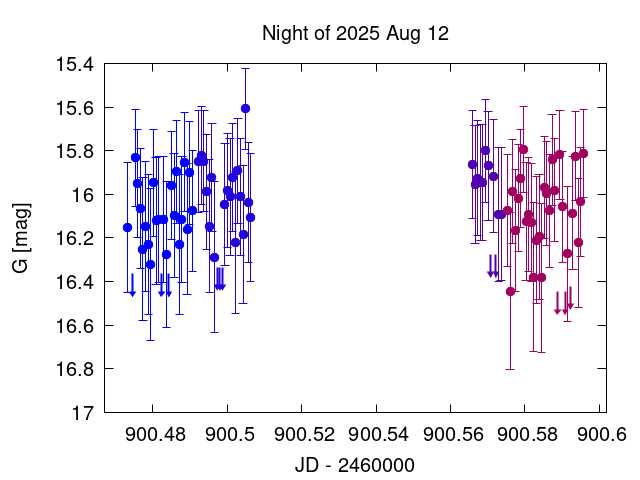}}
\subfigure[
    	\label{fig:unbinmaglc6}]{\includegraphics[width=0.32\textwidth]{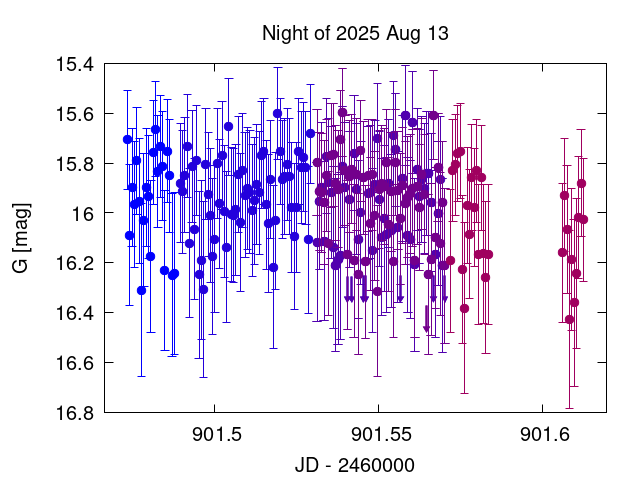}}
\subfigure[
    	\label{fig:unbinmaglc7}]{\includegraphics[width=0.32\textwidth]{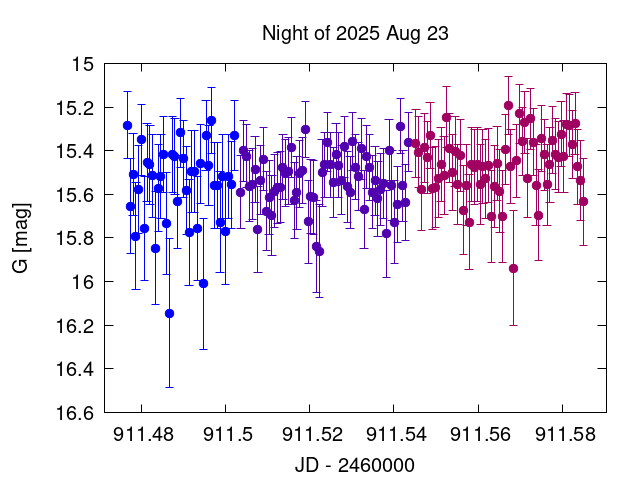}}
\subfigure[
    	\label{fig:unbinmaglc8}]{\includegraphics[width=0.32\textwidth]{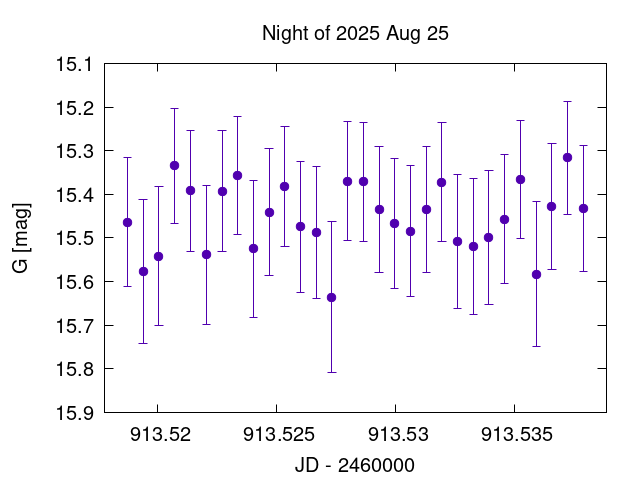}}
\subfigure[
    	\label{fig:unbinmaglc9}]{\includegraphics[width=0.32\textwidth]{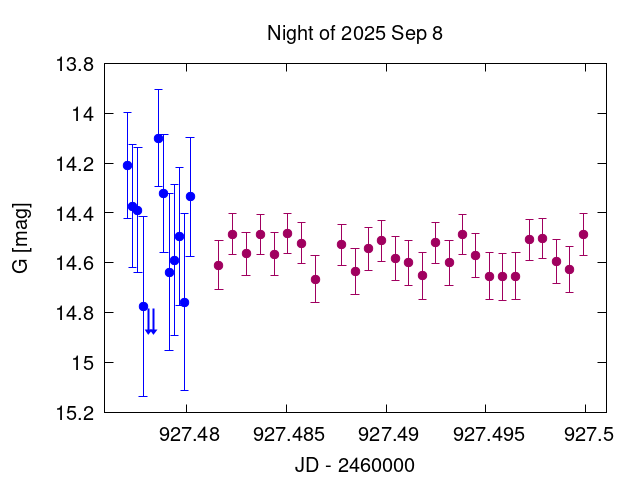}}
\subfigure[
    	\label{fig:unbinmaglc10}]{\includegraphics[width=0.32\textwidth]{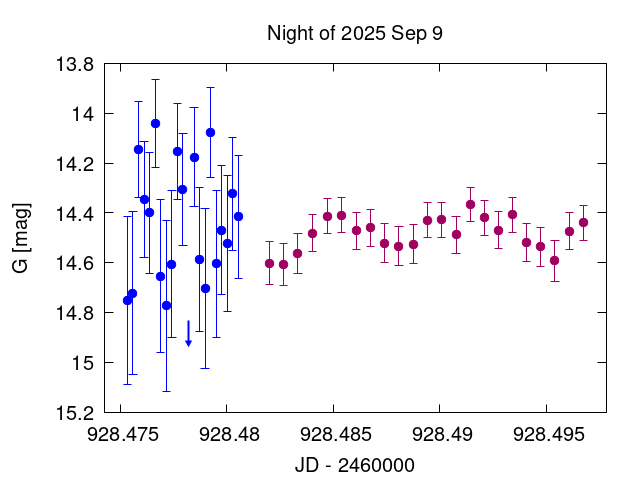}}
\subfigure[
    	\label{fig:unbinmaglc11}]{\includegraphics[width=0.32\textwidth]{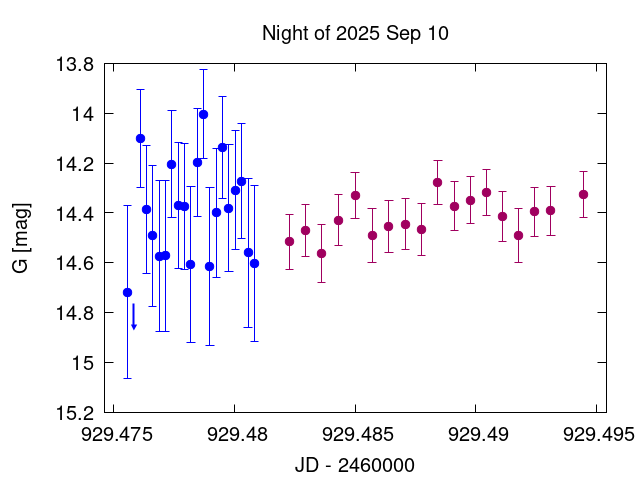}}
\subfigure[
    	\label{fig:unbinmaglc12}]{\includegraphics[width=0.32\textwidth]{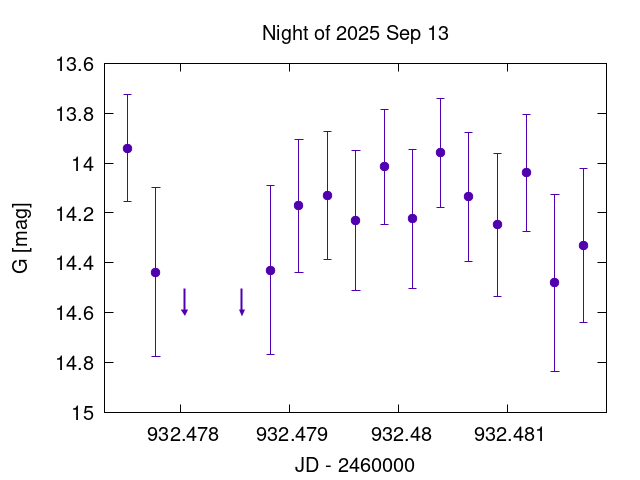}}
\caption{
Un-binned 45\,s cadence clean light curves of 3I/ATLAS from HATPI for 12 different nights, starting from 2025 Aug 6, the first night on which 3I/ATLAS is detectable with $3\sigma$ confidence in individual HATPI exposures. Different colors indicate different IHU/Fields used to observe the target. Arrows show $3\sigma$ upper limits on the magnitude. No variability at the $\sim 0.2$\,mag level is detected within any individual night. We do not show 2025 Sep 11, on which only 4 observations were gathered.
\label{fig:unbinnedmaglc}
}
\end{center}
\end{figure*}

To check for variability on shorter time-scales, Fig.~\ref{fig:unbinnedmaglc} shows the individual, unbinned, photometric light curves from 12 different nights. The first night that we show is 2025 Aug 6, which is the first night on which 3I/ATLAS is detected with $3\sigma$ confidence in individual 45\,s HATPI exposures. We do not see any clear evidence of variability at the $\sim 0.2$\,mag level within any individual night. By Aug 6 the coma of 3I/ATLAS dominates the light from the system, suppressing any rotational signal from the nucleus. A $\sim 0.3$\,mag peak-to-peak amplitude, $P = 16.16 \pm 0.01$\,h periodic variation has been reported based on observations obtained in 2025 July \citep{santanaros:2025}. At this earlier time, the nucleus was still detectable.

\section{Discussion}\label{sec:discussion}

The pre-perihelion light curve of 3I/ATLAS can be used to measure the heliocentric index $n$ of the comet (the dust production as the function of heliocentric distance, assuming a dependence of $\sim r_H^{-n}$), which in turn depends on the composition of the sublimating material. \citet{jewitt:2025} combined photometry of 3I/ATLAS from the Nordic Optical Telescope, together with previously published light curves from the Zwicky Transient Facility \citep[ZTF;][]{ye:2025}, ATLAS \citep{tonry:2025}, and TESS \citep{martinez:2025} to measure an index of $n = 3.8 \pm 0.3$ over a distance range of 4.6\,AU to 1.8\,AU. This index is consistent with CO$_{2}$ being the dominant sublimating material, as independently indicated by initial spectroscopic observations of the tail \citep{lisse:2025a,cordiner:2025}. 

We repeat this analysis, incorporating the HATPI observations, as well as spectrophotometric time-series observations made with the SuperNova Integral Field Spectrograph (SNIFS) on the University of Hawai$^{\text{`}}$i 2.2\,m telescope \citep{hoogendam:2025b}. Following Eq.~1 from \citet{jewitt:2025}, the apparent magnitude of the comet in filter $F$ at a heliocentric distance $r_{H}$ and geocentric distance $\Delta$ is given by
\begin{equation}
    m_{F} = M_{F} + 2.5\log_{10}(r_{H}^n\Delta^{m}) - 2.5\log_{10}(\Phi_{\alpha}).
    \label{eqn:mf}
\end{equation}
Here $\Phi$ is the phase function of the comet, which depends on the Sun-object-Earth angle $\alpha$. $M_{F}$ is the absolute magnitude of the comet through filter $F$, which corresponds to the apparent magnitude that would be observed from a geocentric distance of $1$\,AU when the comet is at a heliocentric distance of $1$\,AU, and with $\Phi = 1$. The heliocentric and geocentric indices of the comet are $n$ and $m$, respectively.

\begin{deluxetable}{lrr}
\tabletypesize{\scriptsize}
\tablecaption{
    Parameters from fitting eq.~\ref{eqn:mf} to the HATPI and literature light curves.
    \label{tab:params}
}
\tablehead{
    Parameter &
    Value &
    Extra Error\tablenotemark{a}
}
\startdata
$n$ & $\bestfitn$ & \\
$\beta$ (mag\,deg$^{-1}$) & $\bestfitbeta$ & \\
$M_{HATPI,G}$ (mag) & $\bestfitMG$ & $\extraerrorG$ \\
$M_{NOT,R}$ (mag) & $\bestfitMR$ & $\extraerrorR$ \\
$M_{ZTF,g}$ (mag) & $\bestfitMgz$ & $\extraerrorgz$ \\
$M_{ZTF,r}$ (mag) & $\bestfitMrz$ & $\extraerrorrz$ \\
$M_{ATLAS,c}$ (mag) & $\bestfitMc$ & $\extraerrorc$ \\
$M_{ATLAS,o}$ (mag) & $\bestfitMo$ & $\extraerroro$ \\
$M_{ATLAS,w}$ (mag) & $\bestfitMw$ & $\extraerrorw$ \\
$M_{TESS,T}$ (mag) & $\bestfitMT$ & $\extraerrorT$ \\
$M_{SNIFS,g}$ (mag) & $\bestfitMhg$ & $\extraerrorhg$ \\
$M_{SNIFS,c}$ (mag) & $\bestfitMhc$ & $\extraerrorhc$ \\
$M_{SNIFS,r}$ (mag) & $\bestfitMhr$ & $\extraerrorhr$ \\
$M_{SNIFS,o}$ (mag) & $\bestfitMho$ & $\extraerrorho$ \\
$M_{SNIFS,i}$ (mag) & $\bestfitMhi$ & $\extraerrorhi$ \\
$M_{SNIFS,z}$ (mag) & $\bestfitMhz$ & $\extraerrorhz$ \\
\enddata
\tablenotemark{a}{Extra error terms added in quadrature to the uncertainties for each light curve so that the best-fit model has $\chi^{2}$ per degree of freedom equal to 1 for each light curve.}
\end{deluxetable}

\begin{deluxetable*}{rrrrrrrrrrrrrrrr}
\tablewidth{0pc}
\tabletypesize{\scriptsize}
\tablecaption{
    Nightly Binned and Cleaned HATPI Light Curve of 3I/ATLAS shown in Fig.~\ref{fig:fitindex}.
    \label{tab:nightlybinnedmaglc}
}
\tablehead{
    \multicolumn{1}{c}{JD} &
    \multicolumn{1}{c}{Gmag$_0$\tablenotemark{a}} &
    \multicolumn{1}{c}{eGmag$_0$\tablenotemark{a}} &
    \multicolumn{1}{c}{Gmag$_1$\tablenotemark{b}} &
    \multicolumn{1}{c}{eGmag$_1$\tablenotemark{b}} &
    \multicolumn{1}{c}{Gmag$_2$\tablenotemark{c}} &
    \multicolumn{1}{c}{eGmag$_2$\tablenotemark{c}} &
    \multicolumn{1}{c}{Ap. Size\tablenotemark{d}} &
    \multicolumn{1}{c}
    {Gmag} &
    \multicolumn{1}{c}{eGmag} &
    \multicolumn{1}{c}{$r_{h}$\tablenotemark{e}} &
    \multicolumn{1}{c}{$\Delta$\tablenotemark{f}} &
    \multicolumn{1}{c}{$\alpha$\tablenotemark{g}} &
    \multicolumn{1}{c}{$\nu$\tablenotemark{h}} &
    \multicolumn{1}{c}{Nobs\tablenotemark{i}} &
    \multicolumn{1}{c}
    {$\Delta$t\tablenotemark{j}}\\
    \multicolumn{1}{c}{(d)} &
    \multicolumn{1}{c}{(mag)} &
    \multicolumn{1}{c}{(mag)} &
    \multicolumn{1}{c}{(mag)} &
    \multicolumn{1}{c}{(mag)} &
    \multicolumn{1}{c}{(mag)} &
    \multicolumn{1}{c}{(mag)} &
    \multicolumn{1}{c}{(arcsec)} &
    \multicolumn{1}{c}{$7 \times 10^{4}$ km} &
    \multicolumn{1}{c}{$7 \times 10^{4}$ km} &
    \multicolumn{1}{c}{(AU)} &
    \multicolumn{1}{c}{(AU)} &
    \multicolumn{1}{c}{(deg)} &
    \multicolumn{1}{c}{(deg)} &
    \multicolumn{1}{c}{} &
    \multicolumn{1}{c}{(h)} \\
    \multicolumn{1}{c}{} &
    \multicolumn{1}{c}{} &
    \multicolumn{1}{c}{} &
    \multicolumn{1}{c}{} &
    \multicolumn{1}{c}{} &
    \multicolumn{1}{c}{} &
    \multicolumn{1}{c}{} &
    \multicolumn{1}{c}{} &
    \multicolumn{1}{c}{(mag)} &
    \multicolumn{1}{c}{(mag)} &
    \multicolumn{1}{c}{} &
    \multicolumn{1}{c}{} &
    \multicolumn{1}{c}{} &
    \multicolumn{1}{c}{} &
    \multicolumn{1}{c}{} &
    \multicolumn{1}{c}{}
}
\startdata
2460859.69823 & 17.797 & 0.082 & 17.716 & 0.091 & 17.66  & 0.10  & 28.6 & 17.796 & 0.082 & 4.441 & 3.440 & 2.5954 & 281.0854  & 281 & 6.36 \\
2460860.64642 & 17.682 & 0.075 & 17.791 & 0.098 & 17.92  & 0.13  & 28.8 & 17.684 & 0.074 & 4.410 & 3.412 & 2.9015 & 281.2339 & 284 & 5.39 \\
2460861.70327 & 17.253 & 0.049 & 17.198 & 0.054 & 17.334 & 0.069 & 29.1 & 17.251 & 0.046 & 4.375 & 3.382 & 3.2539 & 281.4020 & 381 & 6.34 \\
2460873.66564 & 17.248 & 0.086 & 17.28  & 0.10  & 17.38  & 0.13  & 32.0 & 17.260 & 0.067 & 3.978 & 3.078 & 7.7323 & 283.5111 & 93 & 1.40 \\
2460874.50012 & 17.040 & 0.054 & 17.086 & 0.069 & 17.100 & 0.081 & 32.2 & 17.057 & 0.042 & 3.950 & 3.060 & 8.0670 & 283.6739 & 200 & 5.59 \\
2460880.56973 & 16.911 & 0.064 & 16.873 & 0.072 & 16.920 & 0.088 & 33.5 & 16.892 & 0.048 & 3.751 & 2.939 & 10.5501 & 284.9306 & 97 & 3.28 \\
2460881.47923 & 16.782 & 0.092 & 16.73  & 0.11  & 16.76  & 0.12  & 33.7 & 16.755 & 0.073 & 3.721 & 2.923 & 10.9254 & 285.1304 & 36 & 5.49 \\
2460882.58583 & 16.924 & 0.048 & 16.922 & 0.061 & 16.972 & 0.073 & 33.9 & 16.923 & 0.039 & 3.685 & 2.903 & 11.3830 & 285.3781 & 181 & 5.29 \\
2460894.60229 & 16.396 & 0.032 & 16.396 & 0.038 & 16.422 & 0.043 & 36.0 & 16.396 & 0.029 & 3.294 & 2.735 & 16.2039 & 288.4154 & 301 & 3.84 \\
2460895.59798 & 16.289 & 0.033 & 16.287 & 0.039 & 16.320 & 0.045 & 36.1 & 16.287 & 0.031 & 3.262 & 2.724 & 16.5787 & 288.6995 & 279 & 3.76 \\
2460896.54977 & 16.137 & 0.034 & 16.064 & 0.036 & 16.023 & 0.042 & 36.3 & 16.080 & 0.029 & 3.232 & 2.715 & 16.9312 & 288.9764 & 201 & 3.70 \\
2460899.50209 & 16.044 & 0.019 & 16.015 & 0.022 & 16.006 & 0.026 & 36.6 & 16.020 & 0.018 & 3.137 & 2.687 & 17.9877 & 289.8696 & 225 & 2.65 \\
2460900.50528 & 16.076 & 0.029 & 16.032 & 0.034 & 16.009 & 0.041 & 36.8 & 16.039 & 0.029 & 3.105 & 2.678 & 18.3323 & 290.1854 & 88 & 2.94 \\
2460901.54540 & 15.964 & 0.015 & 15.930 & 0.017 & 15.899 & 0.020 & 36.9 & 15.935 & 0.015 & 3.072 & 2.669 & 18.6813 & 290.5198 & 305 & 3.34 \\
2460911.53092 & 15.508 & 0.013 & 15.451 & 0.015 & 15.415 & 0.018 & 37.8 & 15.455 & 0.015 & 2.759 & 2.606 & 21.4902 & 294.1352 & 161 & 2.60 \\
2460913.52832 & 15.450 & 0.027 & 15.382 & 0.031 & 15.351 & 0.037 & 37.9 & 15.386 & 0.030 & 2.697 & 2.597 & 21.9124 & 294.9587 & 30 & 0.46 \\
2460927.48611 & 14.561 & 0.016 & 14.453 & 0.015 & 14.374 & 0.016 & 38.5 & 14.452 & 0.015 & 2.280 & 2.555 & 23.1730 & 301.9335 & 40 & 0.55 \\
2460928.48232 & 14.479 & 0.015 & 14.371 & 0.014 & 14.294 & 0.015 & 38.6 & 14.370 & 0.014 & 2.251 & 2.553 & 23.1333 & 302.5287 & 44 & 0.51 \\
2460929.48056 & 14.411 & 0.022 & 14.286 & 0.019 & 14.180 & 0.023 & 38.6 & 14.284 & 0.018 & 2.223 & 2.550 & 23.0739 & 303.1406 & 39 & 0.45 \\
2460930.49017 & 14.273 & 0.043 & 14.120 & 0.034 & 13.990 & 0.030 & 38.6 & 14.116 & 0.033 & 2.194 & 2.548 & 22.9935 & 303.7754 & 4 & 0.05 \\
2460932.47974 & 14.246 & 0.071 & 14.076 & 0.076 & 13.935 & 0.082 & 38.7 & 14.071 & 0.073 & 2.139 & 2.544 & 22.7737 & 305.0758 & 16 & 0.10 \\
\enddata
\tablenotetext{a}{Using 28.6\arcsec\ aperture.}
\tablenotetext{b}{Using 38.4\arcsec\ aperture.}
\tablenotetext{c}{Using 46.3\arcsec\ aperture.}
\tablenotetext{d}{Size in arcseconds of the $7 \times 10^{4}$\,km linear aperture.}
\tablenotetext{e}{Heliocentric distance of 3I/ATLAS.}
\tablenotetext{f}{Geocentric distance of 3I/ATLAS.}
\tablenotetext{g}{Sun-3I/ATLAS-Earth angle.}
\tablenotetext{h}{Apparent true anomaly of 3I/ATLAS in its heliocentric orbit.}
\tablenotetext{i}{Number of individual photometric measurements contributing to the binned value.}
\tablenotetext{j}{The time-span of the individual observations that are binned together.}
\end{deluxetable*}

As noted by \citet{jewitt:2025}, one can fix $m \equiv 2$ when using a fixed linear aperture to measure photometry (i.e., an aperture of angular size that varies with the distance to the comet $\Delta$ as $\Delta^{-1}$, so that it subtends a fixed linear span at the distance of the comet), but when using an aperture of fixed angular size $m$ will in general take a different value. These authors adopted a fixed linear aperture of $10^4$\,km. Due to the lower spatial resolution of HATPI, we adopt a larger linear aperture of $7 \times 10^4$\,km, such that the minimum angular aperture of 28.6\arcsec\ is used at the first clean HATPI observation, when $\Delta$ is maximum among the HATPI observations. To estimate the photometry within the fixed linear aperture at each observation, we perform linear interpolation, in calibrated fluxes, between the three different apertures used. The resulting binned data are given in Table~\ref{tab:nightlybinnedmaglc}.

\begin{figure*}[t]
\begin{center}
\includegraphics[width=2\columnwidth]{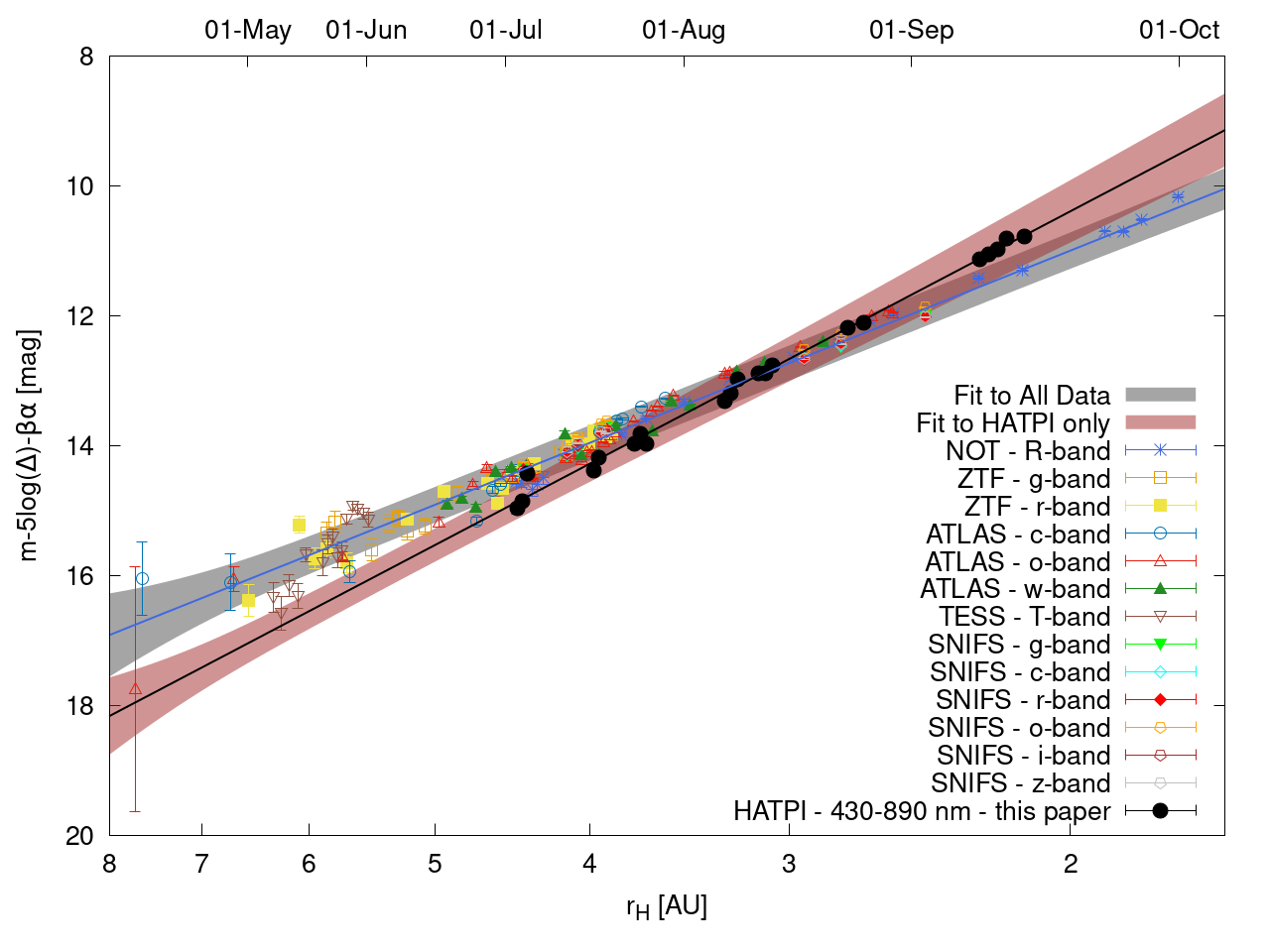}
\caption{
A comparison of the nightly-binned clean $7 \times 10^{4}$\,km linear-aperture HATPI light curve of 3I/ATLAS to light curves reported in the literature. The ATLAS photometry is taken from \citet{tonry:2025}, the ZTF photometry from \citet{ye:2025}, the TESS photometry from \citet{martinez:2025}, and the NOT photometry from \citet{jewitt:2025}. We plot the observed magnitude, corrected for the geocentric distance, $\Delta$, and the phase-function $\beta\alpha$, with $\beta$ fixed to its optimized value of $\bestfitbeta$\,mag\,deg$^{-1}$ for the plot. We plot this as a function of the heliocentric distance $r_{H}$, shown on a logarithmic scale. All light curves have been shifted by $M_{G}-M_{F}$ to match the optimized HATPI heliocentric magnitude. We show the best-fit power-law model (Eq.~\ref{eqn:mf}) when jointly fitting all of the observations, together with the model from fitting only the HATPI observations, while using a prior to constrain $\beta$ to its jointly optimized value. The bands about each best-fit line show the $1\sigma$ confidence interval for the model. The HATPI observations show a steeper increase in brightness with decreasing $r_{H}$, with a power-law index of $n = \bestfitnhatpionly$ compared to $n = \bestfitn$ when fitting all of the data.
\label{fig:fitindex}
}
\end{center}
\end{figure*}

Fig.~\ref{fig:fitindex} compares the fixed linear aperture HATPI photometry to photometry from the literature, using the same datasets as \citet{jewitt:2025}, as well as the SNIFS data from \citet{hoogendam:2025b}. In this figure, we also show the result of fitting Eq.~\ref{eqn:mf} to the observations.  In performing this fit we follow \citet{jewitt:2025} in assuming $-2.5\log_{10}(\Phi(\alpha)) = \beta\alpha$. We jointly fit all observations together, optimizing $M_{F}$ for each dataset, together with $n$ and $\beta$. We performed the fit iteratively, adding a separate error term in quadrature to the uncertainties from each light curve until we achieved $\chi^{2}$ per degree of freedom equal to unity for each light curve. The best fit parameters, together with the extra error terms added to each light curve, are listed in Table~\ref{tab:params}. We find $n = \bestfitn$ and $\beta = \bestfitbeta$\,mag\,deg$^{-1}$. The final fit has $\chi^2 = 206.9$. There are 218 observations and 16 parameters that are varied, so that the combined reduced $\chi^2$ is $1.02$.

The HATPI data show a steeper increase in brightness with decreasing $r_{H}$ when compared to the literature observations. The difference is mostly seen between the HATPI and NOT measurements, which extend to the closest heliocentric separations. When we fit the HATPI data alone, applying a prior on $\beta$ of $\beta = \bestfitbeta$\,mag\,deg$^{-1}$ based on the fit to all of the observations, we find $M_{G} = \bestfitMGhatpionly$\,mag and $n = \bestfitnhatpionly$. In this case an extra error of \extraerrorGhatpionly\,mag is added in quadrature to the HATPI light curve uncertainties to achieve a reduced $\chi^2$ of unity.

A difference in the heliocentric index for HATPI is not unexpected given the larger linear aperture used for the HATPI photometry ($7 \times 10^{4}$\,km for HATPI, compared with $10^{4}$\,km for NOT), together with the different band-pass of HATPI (430\,nm to 890\,nm for HATPI, compared with 568\,nm to 718\,nm for the NOT $R$-band).  

The bluer wavelength range of HATPI may be picking up gas emission in addition to scattering of optical light by dust in the coma. Pre-perihelion spectroscopic observations by \citet{hoogendam:2025b}, covering the time period up to 2025 Sep 2, show Ni, Fe and CN emission lines. The reddest of these are the CN lines around 387\,nm, which are outside the blue cutoff of the HATPI band-pass. Similarly \citet{salazarmanzano:2025} detect the CN lines, in observations made in 2025 Aug, but do not detect redder emission that might come from C$_3$ or C$_2$ lines. Post-perihelion observations by \citet{hoogendam:2026} do show C$_2$ emission lines that are within the HATPI band-pass. It is unclear, however, whether such emission could have been present during the latest HATPI observations obtained before peri-helion, between 2025 Sep 8 and 13.

Because HATPI has a lower spatial resolution than ATLAS, NOT, ZTF or SNIFS, we are forced to use a larger photometric aperture that picks up more of the extended tail of the comet. Differences between the extended tail of the comet and the inner coma (e.g., in the dust composition, scattering properties, gas emission, or in the phase dependence) could also be responsible for the steeper heliocentric index seen in the HATPI observations. A hint of this can be seen by adopting a larger linear aperture of $8 \times 10^{4}$\,km, such that the largest angular-size aperture is used when 3I/ATLAS achieves the smallest value of $\Delta$ within the HATPI observations. Using the $8 \times 10^{4}$\,km aperture leads to a slightly steeper heliocentric index of $n = 5.31 \pm 0.11$ compared to $n = \bestfitnhatpionly$ when using the $7 \times 10^{4}$\,km aperture.

We also note that a steeper heliocentric index is expected for H$_{2}$O sublimation, and spectroscopic observations by SPHEREx in early August have revealed significant water gas emission in addition to CO$_2$ emission \citep{lisse:2025b}. H$_2$O has also been seen in post-perihelion observations by \citet{lisse:2026}.

We also note that comets typically change color as they approach perihelion, becoming bluer over time. \citet{tonry:2025} noted a change in color for 3I/ATLAS, which they observed to become bluer after 2025 July 13. This could also manifest as a steeper power-law index in a bluer filter, though the synthetic SNIFS $g$-band photometry from \citet{hoogendam:2025b} appears to be well-fit by the same power-law used to fit all of the observations. Our assumption of a fixed {\em Gaia} color when converting the HATPI measurements to {\em Gaia} $G$-band is also inaccurate in this sense, and could lead to a $\sim 0.02$\,mag systematic error over the course of the observations. Such a shift is too small to be responsible for the steeper rise in brightness seen in the HATPI observations.

\section{Conclusion}\label{sec:conclusion}

In this paper we have reported time-series photometry observations of the interstellar comet 3I/ATLAS obtained with the HATPI instrument. This is the first report of moving object time-series photometry from the HATPI facility, and we describe the methods we have used to extract these measurements, together with our process for identifying clean photometry measurements.

A total of 7294 clean 45\,s photometric measurements of 3I/ATLAS were obtained over 46 nights between 2025 May 2 and 2025 Sep 13, inclusive. After binning all observations obtained on a given night, we find that HATPI first robustly recovers 3I/ATLAS on the night of 2025 July 2, at a {\em Gaia} $G$-band magnitude of $17.796 \pm 0.082$\,mag. The earliest night on which 3I/ATLAS is detected in individual 45\,s exposures is 2025 Aug 6. We do not see any clear intra-night variability at short time-scales above $\sim 0.2$\,mag after this date.

The nightly binned HATPI observations are well-fit as a power-law function of the heliocentric distance, with a heliocentric index of $n = \bestfitnhatpionly$. This is steeper than the dependence seen in other light curves in the literature. A simultaneous fit to the observations from HATPI, NOT, ATLAS, ZTF, TESS, and SNIFS yields a heliocentric index of $n = \bestfitn$, and a phase function of $\beta = \bestfitbeta$\,mag\,day$^{-1}$. The steeper dependence of HATPI may be due to the larger photometric aperture for HATPI that picks up more of the extended tail of the comet. It is also possible that the HATPI band-pass includes gas emission lines, though emission lines within the HATPI band-pass were not detected in pre-perihilion spectra of 3I/ATLAS.

Thanks to its massive field-of-view, and continuous high-cadence monitoring, numerous moving objects have been observed by HATPI already, and will be observed in the future. The processes that we have outlined in this paper can be applied to these objects as well, including many objects that have not received the same degree of intensive targeted observation as 3I/ATLAS.  

\begin{acknowledgements}

We thank the anonymous referee for helpful comments that have improved the quality of this work. Funding for developing and prototyping the HATPI system and initial construction was provided by the David and Lucile Packard Foundation.
Funding for the construction and operation of HATPI has been provided by the Gordon and Betty Moore Foundation. The Mt.\ Cuba Astronomical Foundation funded the construction of the HATPI mount. Data presented here is based on observations obtained with HATPI, stationed at the Las Campanas Observatory of Carnegie Science. 
We thank the Observatory, staff and management, for hosting HATPI at this world-class facility, and providing support for its operations. 
We thank the Jet Propulsion Laboratory (JPL) for providing access to the HORIZONS ephemeris computation service for generating solar system object trajectories.
AJ acknowledges support from Fondecyt project 1251439. 
GJT acknowledges support from the UK Space Agency as part of its support for the UK component of the PLATO Mission.

\end{acknowledgements}

\facilities{Gaia, HATPI, NOT, ZTF, TESS, ATLAS, SNIFS}

\software{FITSH \citep{pal:2012}, Astropy \citep{astropy:2013,astropy:2018,astropy:2022}, VARTOOLS \citep{hartman:2016:vartools}}


\bibliography{hatpibib}
\bibliographystyle{aasjournal}

\end{document}